\documentclass[aps,superscriptaddress,tightenlines,eqsecnum,nofootinbib]{revtex4}
\usepackage{graphicx,amssymb,amsbsy,bm,amsmath}

\usepackage{hyperref}

\newcommand{\Ok}{\ensuremath{\Omega_k}}
\newcommand{\Okr}{\ensuremath{\Omega^R_k}}
\newcommand{\vx}{\ensuremath{\vec{x}}}
\newcommand{\vy}{\ensuremath{\vec{y}}}
\newcommand{\vk}{\ensuremath{\vec{k}}}
\newcommand{\vq}{\ensuremath{\vec{q}}}

\newcommand{\be}{\begin{equation}}
\newcommand{\ee}{\end{equation}}
\newcommand{\bea}{\begin{eqnarray}}
\newcommand{\eea}{\end{eqnarray}}

\begin{document}
\title{Effective Field Theory out of Equilibrium: \\ Brownian quantum fields.}
\author{D. Boyanovsky}
\email{boyan@pitt.edu} \affiliation{Department
of Physics and Astronomy,\\ University of Pittsburgh\\Pittsburgh,
Pennsylvania 15260, USA}
\date{\today}
\begin{abstract}
The emergence of an effective field theory out of equilibrium is studied in the case in which a light field --the system-- interacts with very heavy  fields in a finite temperature bath.   We obtain the reduced density matrix for the light field, its time evolution is determined by an effective action that includes the \emph{influence action} from correlations of the heavy degrees of freedom.   The non-equilibrium effective field theory yields a Langevin equation of motion for the light field in terms of dissipative and noise kernels that obey a generalized fluctuation dissipation relation.   These are completely determined by the spectral density of the bath which is analyzed in detail for several cases. At $T=0$ we elucidate the effect of thresholds in the renormalization aspects and the asymptotic emergence of a local effective field theory with unitary time evolution. At $T\neq 0$ new ``anomalous'' thresholds arise, in particular the \emph{decay} of the environmental heavy fields   into the light field leads to   \emph{dissipative} dynamics of  the light field. Even when the heavy bath particles are thermally suppressed this dissipative contribution leads to the \emph{thermalization} of the light field which is  confirmed by a quantum kinetics analysis. We obtain the quantum master equation and show explicitly that its solution in the field basis is precisely the influence action that determines the effective non-equilibrium field theory. The Lindblad form of the quantum master equation features \emph{time dependent dissipative coefficients}. Their time dependence is crucial to extract renormalization effects  at asymptotically   long time. The dynamics from the quantum master equation is in complete agreement with that of the effective  action, Langevin dynamics and quantum kinetics, thus providing a  unified framework to    effective field theory out of equilibrium.
\end{abstract}


\maketitle

\section{Introduction}\label{sec:intro}
Effective field theory is a powerful organizational principle to describe phenomena below some energy scale, or alternatively on large spatio-temporal scales, and is ubiquitous across fields. Several applications of effective field theory have become the pillars of fundamentally important paradigms, for example: universality in critical phenomena emerges at long wavelengths after coarse graining over short wavelength degrees of freedom a l\'a \emph{Wilson}:  the Landau-Ginsburg theory of phase transitions classifies universality classes in terms of few coarse grained order parameters and their symmetries\cite{ma},
 the Landau-Ginsburg theory of superconductivity emerges after integrating out the fermionic quasiparticles leading to an effective theory of the superconducting  order parameter\cite{fradkin}, and  hydrodynamics   is a description of long wavelength collective flow that emerges  after coarse graining over small scales and is valid on scales much larger than a microscopic mean free path. These are but a few historically and conceptually important examples of effective field theories. In particle physics\cite{eft1,eft2,eft3,eft4} effective field theory provides a systematic characterization of   phenomena below some energy scale and is an important tool to incorporate the physics associated with the degrees of freedom beyond such energy scale in a consistent and systematic manner. These concepts have been extended to the realm of early Universe cosmology to describe cosmological perturbations\cite{weinberg,cheung}, large scale structure formation\cite{senatore} and could ultimately    underlie the description    of inflationary cosmology in terms of an inflaton scalar field as an effective dynamical degree of freedom well below the Planck scale\cite{nuestroreviu}. The usual approach to effective field theory begins by recognizing the operators that could enter in an effective Lagrangian based on internal and space-time symmetries in a suitable derivative expansion, where higher derivatives and higher dimensional operators are suppressed by inverse powers of the (high) energy scale\cite{eft1,eft2,eft3,eft4}, such a description is manifestly local and yields unitary time evolution of observables.
 However, it is \emph{not} always the case that integrating over ``hard'' scales the resulting low energy effective field theory is local, an important counterexample is the ``hard thermal loop'' effective field theory at finite temperature\cite{htl,lebellac}. At finite temperature, Landau damping is a medium process that yields spectral densities in loop diagrams that feature support below the light cone and leads to non-local terms in the effective action\cite{htl}  and long time tails in the dynamics\cite{boylan}.
 Fundamentally, effective field theory must be understood as emerging from ``tracing over'', ``integrating out'' or ``coarse graining''  high energy degrees of freedom with fast dynamics and short wavelength fluctuations, and it describes the  \emph{influence}  of the high energy degrees of freedom  over the low energy (slow, long wavelength) degrees of freedom. Cast in this manner, effective field theory is another manifestation of the quantum open system approach to  studying the effect of an environment upon the dynamics of a system\cite{breuer,zoeller,weiss,daley} pioneered with the study of quantum Brownian motion\cite{feyver,schwinger,leggett}. This approach begins by considering the time evolution of the full density matrix of the system coupled to the environment and tracing over the environmental (or ``bath'') degrees of freedom leading to a \emph{reduced} density matrix for the system, whose time evolution includes the \emph{influence action} from the environment\cite{feyver} upon the system. Within this context the effective action arises as the sum of the system's action and the influence action. The effective equations of motion for the system variables become a quantum Langevin equation with a dissipative term and stochastic force that obey  a generalized fluctuation dissipation relation\cite{ford,schmid,grabert}. Quantum Brownian motion of a particle coupled linearly and non-linearly to general environments has been studied thoroughly in refs.\cite{hupaz} and a comprehensive and in-depth discussion of non-equilibrium phenomena and in particular generalizations of quantum Brownian motion is available in refs.\cite{calhubuk,calhu,flemhu}. Decoherence and effective stochastic dynamics emerging from tracing over short wavelength degrees of freedom  are also of fundamental importance in cosmology\cite{staro,calhuuniv,stoca,proko1,woodard1,prokowood,onemli}.

 The influence action approach has recently been argued to provide an effective   description of the dynamics of long-wavelength fluctuations when combined with a Wilsonian approach to coarse graining the short wavelength components\cite{bala} which are taken as an environment or bath.  An alternative approach to quantum open systems relies on the quantum master equation\cite{breuer,zoeller,weiss,daley} for the reduced density matrix. This approach has recently been advocated in   cosmology\cite{burhol} under the  assumption  that the environment only features short time (delta function) correlations.

\vspace{1mm}

\textbf{Goals :} In this article we study the emergence of an effective field theory description out of equilibrium in the case in which a light field --the system--   interacts  with very heavy fields, taken to be the ``environment'' and integrated out or traced over. In particular we focus on  analyzing in detail the influence of correlations of the heavy degrees of freedom, upon the non-equilibrium dynamics of the light degrees of freedom. We seek to elucidate in a direct manner the relationship between the influence action, stochastic  and quantum master equation approaches without \emph{assumptions} on environmental correlations, focusing precisely on how the spectral properties of the environment lead to the different dynamical time scales of the low energy degrees of freedom. We  address  the following questions: \textbf{(I):} In quantum field theory  there are thresholds to excitation of heavy degrees of freedom, a local and unitary effective action emerges from integrating out heavy (or high energy) degrees of freedom when the energy and momentum associated with the dynamics of the light fields is well below this threshold. How are these threshold effects manifest in the influence action, stochastic and quantum master equation description?. \textbf{(II):} When the heavy degrees of freedom form a thermalized plasma through their mutual interaction, there are new ``in medium'' corrections to the spectral density of correlators of the heavy degrees of freedom that \emph{may} lead to dissipative processes in the influence action of the light degrees of freedom even when the $T=0$ thresholds correspond to high energy. This is the case in ``hard thermal loops''\cite{htl,lebellac} as a consequence of Landau damping. Do these new, ``in medium'' contributions to the spectral density lead to dissipative dynamics of the light field?. \textbf{(III):} does the light field thermalize with the heavy coarse grained degrees of freedom?. If the light field thermalizes with the ``bath'' of heavy particles, do the influence functional and quantum master equation approaches agree with the dynamics of thermalization from quantum kinetics?. \textbf{(IV):} Heavy environmental fields may actually \emph{decay} into the lighter species considered as the ``system'' when they are coupled, how does the decay of the environmental degrees of freedom affect the dynamics of the light fields?. \textbf{(V):} How do renormalization aspects emerge from the non-equilibrium descriptions: influence action, stochastic and quantum master equation approaches?.

\vspace{1mm}

\vspace{1mm}

 \textbf{Summary of results:}
 To answer these questions we focus on the case of a light scalar field $\phi$ of mass $m_\phi$ coupled to either just one scalar heavy field $\chi$ with $M\gg m_\phi$ or two heavy scalar fields $\chi_1,\chi_2$ with a hierarchy of masses $M_1> M_2 \gg m_\phi$. The heavy fields generically denoted as $\chi$ are the environmental degrees of freedom, which are integrated out or ``traced over'', the light field $\phi$ constitutes the ``system''. We study   couplings of the  form $J[\phi]~\mathcal{O}[\chi_j]$ where $J[\phi]$ and $\mathcal{O}[\chi_j]$ are generic polynomials in the respective fields. We focus in detail on two types of couplings from which we draw more general conclusions: \textbf{(i):} $g \phi^2 \chi$, \textbf{(ii):} $g \phi \chi_1 \chi_2$ both at $T=0$ and $ T\neq 0$.
 The first type yields an effective ``current-current'' interaction in the low energy limit and illustrates how this familiar local effective field theory emerges in the influence action and quantum master equation approaches in the long time limit. The second type introduces a rich spectral density with ``anomalous'' thresholds to excitation of  the degrees of freedom of the environment as a consequence of in medium effects. As a consequence of the heavy  field  with $M_1$ decaying into the light field $\phi$  this spectral density features support on the mass shell of the light particles leading to dissipative phenomena with a wealth of dynamical scales. The main results are:
 \begin{itemize}
 \item{We obtain the time evolution of the reduced density matrix and the non-equilibrium effective action by tracing over the ``bath'' degrees of freedom to order $g^2$ in terms of the various correlation functions of the ``bath'', which are determined by the spectral density in all cases. For the case $g\phi^2\chi$ (single heavy field), we analyze the transient dynamics and obtain the ``current-current'' effective theory in the long time limit. In the case of a thermal bath of two heavy fields $\chi_1,\chi_2$ with $M_1>M_2 \gg m_\phi$ we find the (one loop) spectral density for $T\neq 0$, it features several thresholds \emph{below} the two particle threshold at $T=0$, one describes Landau damping and the other describes the \emph{decay} of the heavier bath field into the light field$\phi$. These two ``anomalous'' thresholds lead to a dissipative contribution to the $\phi$ effective action, in particular we find that the \emph{decay} $\chi_1 \rightarrow \chi_2 \phi$ leads to a purely dissipative contribution to the effective action which describes the \emph{thermalization} of the light field with the bath.   }

 \item{ From the non-equilibrium influence action we obtain a \emph{semiclassical} stochastic description for the light field, the equation of motion for the non-equilibrium average is a generalized Langevin equation with  a stochastic Gaussian noise term and a  non-local dissipative kernel-- hence $\phi$ is a light  Brownian quantum field--, the noise and dissipative kernels obey a generalized fluctuation-dissipation relation. For the coupling $g\phi \chi_1\chi_2$ the noise is additive but with a colored spectrum that reflects the underlying temporal correlations of the bath, for more general couplings $g~J[\phi]\chi_1\chi_2$ with $J[\phi]\neq \phi$ we find multiplicative noise. For the case $g\phi\chi_1\chi_2$  we obtain the solution of the Langevin equation and obtain the correlation functions of the light field, for $T\neq 0$ we find that the decay $\chi_1 \rightarrow \chi_2 \phi$ leads to the \emph{thermalization} of $\phi$ and obtain the time scale for thermalization.  We show that the dynamics of thermalization obtained from the effective action is in complete agreement with a \emph{quantum kinetic} description of $\phi$ thermalization. As $T\rightarrow 0$ only the high energy two particle threshold remains in the spectral density of the bath, which is responsible for \emph{renormalization effects} such as wave function and mass renormalization. In this case a local and unitary effective field theory emerges in the asymptotic long time limit.   }

     \item{We obtain the quantum master equation up to $\mathcal{O}(g^2)$ and show that its solution in the field basis is precisely the influence function obtained by tracing over the bath. In the case of linear coupling to the bath we obtain the Lindblad form of the master equation under clearly specified approximations. The Lindblad form contains a Hamiltonian and a dissipative term, both feature time dependent coefficient functions. The time dependence of the dissipative functions is \emph{crucial} to understand renormalization effects. The quantum master equation describes thermalization in complete agreement with   the influence function, the Langevin stochastic description and the quantum kinetic equation. Keeping the time evolution of the dissipative coefficients in the Lindblad form allows to extract the dynamics of the ``dressing'' of the bare states and wave function renormalization in the asymptotic long time limit, again  in complete agreement with the influence function and Langevin descriptions even in the $T=0$ case when only the high energy thresholds are present in the spectral density of the bath. }

 \end{itemize}

\section{The non-equilibrium effective
action}\label{sec:noneLeff}

 We study  the non equilibrium effective action for the case of interacting
 scalar fields, spinor and vector fields may be included, while technically more involved their treatment follows without major conceptual difficulties. We consider a bosonic field $\phi$ of mass $m_\phi$ referred to as the ``system'' and either a single massive scalar field $\chi$  of mass $M_\chi\gg m_\phi$  or two scalar fields collectively denoted by  $\chi_a~;~i=1,2$ with the hierarchy of masses $M_1>M_2 \gg m_\phi$ and a generic interaction between $\phi$ and $\chi_i$. The Lagrangian density is given by
\begin{equation}\label{lagra}
{\cal L}[\phi(x),\chi_a(x)]= {\cal L}_{0,\phi}[\phi(x)]+{\cal
L}_{0,\chi}[\chi_a(x)]-g~J[\phi(x)]~{\cal O}[\chi_a(x)]
\end{equation}
where the ``current'' $J[\phi]$ and ${\cal O}[\chi_a]$ are in general  non-linear function(als) of the respective fields, and ${\mathcal L}_{0,\phi,\chi}$ are the free field
Lagrangian densities for the fields $\phi$ and $\chi_a$, again self-interactions may be included at the expense of technical complications but with no conceptual difficulties.

Specifically we consider the  following   cases:
\begin{itemize}

\item{ Only one field $\chi$ and $J[\phi(x)] = \phi^2(x)$, namely with the interaction $g\,\phi^2(x)\,\chi(x)$. This interaction describes the exchange of a massive ``vector boson'', $\chi$ whereas $J[\phi]=\phi^2$ describes a bilinear ``current'', as in the coupling between gauge bosons and fermionic degrees of freedom. This model allows us to understand the emergence of a ``local Fermi'' theory in the limit where the frequency and  momentum transferred by the current is much smaller than the mass of the (vector) boson.  }

\item{ Two different fields $\chi_{1,2}$ with $M_1 > M_2 \gg m_\phi$ and  $J[\phi(x)] = \phi(x)$, namely with interaction $g\,\phi(x)\,\chi_1(x)\,\chi_2(x)$. This case   allows us to obtain a Langevin  equation of motion for the system field $\phi$ describing its non-equilibrium dynamics as a ``Brownian'' field. This case will also lead to a detailed understanding of thermalization and dissipative processes by interactions with heavy fields. }

\item{Two different fields with the same hierarchy as the previous case but now with $J[\phi(x)] = \phi^2(x)$ and interaction $g\,\phi^2(x)\,\chi_1(x)\,\chi_2(x)$. This case will showcase important renormalization aspects and highlights the limitations of a local description when the ``bath'' or environmental fields $\chi_{1,2}$ form a plasma.  }

\end{itemize}

While we are ultimately interested in obtaining an effective quantum field theory by tracing out ``heavy degrees of freedom'' in cosmology, in this study we focus on Minkowski space time and consider that the fields  $\chi_i$ are treated as
a bath in equilibrium assuming that the bath fields are sufficiently
strongly coupled so as to guarantee their thermal equilibration.
These fields will be ``integrated out'' yielding a reduced density
matrix for the field $\phi$ in terms of an effective real-time
functional, known as  the influence functional\cite{feyver} in the
theory of quantum brownian motion. The reduced density matrix can be
represented by a path integral in terms of the non-equilibrium
effective action that includes the influence functional.  This
method has been used previously to study quantum brownian
motion\cite{feyver,leggett,grabert} and for   studies of
quantum kinetics beyond the Boltzmann equation\cite{boyalamo,yoshimura,boykev}.

Consider the initial density matrix at a time
$t=0$ to be of the form
\begin{equation}
\hat{\rho}(0) = \hat{\rho}_{\phi}(0) \otimes
\hat{\rho}_{\chi}(0) \,.\label{inidensmtx}
\end{equation}

The initial density matrix of the $\chi_a$ fields will be taken to
describe a statistical ensemble in thermal equilibrium at a temperature
$T=1/\beta$, namely

\begin{equation}\label{rhochi}
\hat{\rho}_{\chi}(0) = e^{-\beta\,H_{0\chi}}\,,
\end{equation}

\noindent where $H_{0\chi}(\chi_a)$ is the free field Hamiltonian for the
fields $\chi_a$. We will now refer collectively to the set of
fields $\chi_a$ simply as $\chi$ to avoid cluttering of indices.

The factorization of the initial density matrix is an assumption often explicitly or implicitly made in the literature, it can be relaxed by including initial correlations, we will not consider here this important case, relegating it to future study. The initial factorization entails that
the time evolution of the system can be described as ``switching-on'' the coupling between fields at the initial time. This will result in transient dynamics, however, we will  focus on the long time evolution.

In the field basis the matrix elements of $\hat{\rho}_{\phi}(0)$
are given by
\begin{equation}
\langle \phi |\hat{\rho}_{\phi}(0) | \phi'\rangle =
\rho_{\phi,0}(\phi ,\phi')~~;~~\langle \chi |\hat{\rho}_{\chi}(0) | \chi'\rangle =
\rho_{\chi,0}(\chi ;\chi')\,,
\end{equation} we emphasize that this is a \emph{functional} density matrix as the field has spatial arguments.
The density matrix for $\phi$   represents an initial out of
equilibrium state or ensemble.

  The physical situation described by this initial initial density matrix is that
  of a  field (or fields)  in thermal equilibrium at a temperature
  $T=1/\beta$, namely a heat bath,  which is put in contact with another system, here represented by the field $\phi$.
   Once the system  and bath are put in contact their mutual interaction will
  evolve the initial state out of equilibrium because the initial density matrix does not commute with the total Hamiltonian.

To obtain the effective quantum field theory    out of equilibrium for the light field $\phi$  we will evolve the initial density matrix in time and trace over the ``bath'' degrees of freedom, leading to a reduced density matrix for $\phi$.    Once we obtain the reduced density matrix for the field
$\phi$ we can compute expectation values or correlation functions
of this field.

The time evolution of the initial density matrix is given by

\begin{equation}
\hat{\rho}(t)= U(t)\hat{\rho}(0)U^{-1}(t)\,, \label{rhooft}
\end{equation}
where
\begin{equation}
U(t) = e^{-iHt}\,. \label{unitimeop} \ee
The total Hamiltonian $H$ is given by
\begin{equation}
H=H_{0 \phi} + H_{0 \chi}+H_I(\phi,\chi)~~;~~H_I(\phi,\chi)=g\int d^3x  ~J[\phi(x)]~{\cal O}[\chi(x)]\,, \label{hami}
\end{equation} and $H_{0\phi},H_{0 \chi}$
 are the free field Hamiltonians for the respective fields.

\bea   \rho(\phi_f,\chi_f;\phi'_f,\chi'_f;t) & = &      \langle \phi_f;\chi_f|U(t)\hat{\rho}(0)U^{-1}(t)|\phi'_f;\chi'_f\rangle \nonumber \\
& = & \int D\phi_i D\chi_i D\phi'_i D\chi'_i ~ \langle \phi_f;\chi_f|U(t)|\phi_i;\chi_i\rangle\,\rho_{\phi,0}(\phi_i;\phi'_i)\,\rho_{\chi,0}(\chi_i;\chi'_i)\,
 \langle \phi'_i;\chi'_i|U^{-1}(t)|\phi'_f;\chi'_f\rangle \label{evolrhot}\eea The $\int D\phi$ etc, are functional integrals where the spatial argument has been suppressed. The matrix elements of the time evolution forward and backward can be written as path integrals, namely
 \bea   \langle \phi_f;\chi_f|U(t)|\phi_i;\chi_i\rangle  & = &    \int \mathcal{D}\phi^+ \mathcal{D}\chi^+\, e^{i \int d^4 x \mathcal{L}[\phi^+,\chi^+]}\label{piforward}\\
 \langle \phi'_i;\chi'_i|U^{-1}(t)|\phi'_f;\chi'_f\rangle &  =  &   \int \mathcal{D}\phi^- \mathcal{D}\chi^-\, e^{-i \int d^3 x \mathcal{L}[\phi^-,\chi^-]}\label{piback}
 \eea where we use the shorthand notation
 \be \int d^4 x \equiv \int_0^t dt \int d^3 x \,,\label{d4xdef}\ee
 $ \mathcal{L}[\phi,\chi] $ is given by (\ref{lagra})   and
 the boundary conditions on the path integrals are
  \bea     \phi^+(\vec{x},t=0)=\phi_i(\vec{x})~;~
 \phi^+(\vec{x},t)  &  =  &   \phi_f(\vec{x})\,,\nonumber \\   \chi^+(\vec{x},t=0)=\chi_i(\vec{x})~;~
 \chi^+(\vec{x},t) & = & \chi_f(\vec{x}) \,,\label{piforwardbc}\\
     \phi^-(\vec{x},t=0)=\phi'_i(\vec{x})~;~
 \phi^-(\vec{x},t) &  = &    \phi'_f(\vec{x})\,,\nonumber \\   \chi^-(\vec{x},t=0)=\chi'_i(\vec{x})~;~
 \chi^-(\vec{x},t) & = & \chi'_f(\vec{x}) \,.\label{pibackbc} \\
 \eea

The field variables $\phi^\pm, \chi^\pm$ along the forward ($+$) and backward ($-$) evolution branches are recognized as those necessary for the Schwinger-Keldysh\cite{schwinger,keldysh,maha,calhubuk} closed time path approach to the time evolution of a density matrix.

\subsection{Tracing over the ``bath'' degrees of freedom: reduced density matrix}
 The reduced density matrix for the light field $\phi$ is obtained by tracing over the bath ($\chi_i$) variables, namely

\be \rho^{r}(\phi_f,\phi'_f;t) = \int D\chi_f \,\rho(\phi_f,\chi_f;\phi'_f,\chi_f;t) \,,\label{rhored} \ee we find
\be \rho^{r}(\phi_f,\phi'_f;t) = \int D\phi_i   D\phi'_i  \,  \mathcal{T}[\phi_f,\phi'_f;\phi_i,\phi'_i;t] \,\rho_\phi(\phi_i,\phi'_i;0)\,, \ee
where the time evolution kernel is given by
\be \mathcal{T}[\phi_f,\phi_i;\phi'_f,\phi'_i;t] = {\int} \mathcal{D}\phi^+ \, \int \mathcal{D}\phi^- \, e^{i  \int d^4x \left[\mathcal{L}_0[\phi^+]-\mathcal{L}_0[\phi^-]\right]}\,e^{i\mathcal{F}[J^+;J^-]} \ee with
the following boundary conditions on the forward ($\phi^+$) and backward  ($\phi^-$) path integrals
\bea &  &   \phi^+(\vec{x},t=0)=\phi_i(\vec{x})~;~
 \phi^+(\vec{x},t)  =   \phi_f(\vec{x}) \nonumber \\
&  &   \phi^-(\vec{x},t=0)=\phi'_i(\vec{x})~;~
 \phi^-(\vec{x},t)  =   \phi'_f(\vec{x}) \,.\label{bcfipm}\eea   $\mathcal{F}[J^+;J^-]$ is the \emph{influence action} where $J^\pm \equiv J[\phi^\pm]$ given by
 \be  e^{i\mathcal{F}[J^+;J^-]}   =   \int D\chi_i \int D\chi'_i D\chi_f  \int \mathcal{D}\chi^+ \int \mathcal{D}\chi^- \, e^{i  \int d^4x \left[\mathcal{L}_0[\chi^+]-g J[\phi^+]\mathcal{O}[\chi^+]\right]}~   e^{-i  \int d^4x \left[\mathcal{L}_0[\chi^-]-g J[\phi^-]\mathcal{O}[\chi^-]\right]}\,\rho_{\chi}(\chi_i,\chi'_i;0) \label{inffunc}\ee the boundary conditions on the path integrals are
 \be \chi^+(\vec{x},t=0)=\chi_i(\vec{x})~;~
 \chi^+(\vec{x},t)=\chi_f(\vec{x})~~;~~ \chi^-(\vec{x},t=0)=\chi'_i(\vec{x})~;~
 \chi^-(\vec{x},t)=\chi'_f(\vec{x}) \,. \label{bcchis} \ee

In the above path integral defining the influence action, $J[\phi^\pm]$ acts as an \emph{external source} coupled to the composite operator $\mathcal{O}(\chi)$, therefore, it is straightforward to conclude that
\be e^{i\mathcal{F}[J^+;J^-]} = \mathrm{Tr} \Big[ \mathcal{U}(t;J^+)\,\rho_\chi(0)\,  \mathcal{U}^{-1}(t;J^-) \Big]\,, \label{trasa}\ee where $J^\pm\equiv J[\phi^\pm]$ and $\mathcal{U}(t;J^\pm)$ is the   time evolution operator in the $\chi$ sector in presence of \emph{external sources} $J^\pm$ namely \be \mathcal{U}(t;J^+) = T\Big( e^{-i \int_0^t H_\chi[J^+(t')]dt'}\Big) ~~;~~
\mathcal{U}^{-1}(t;J^-) = \widetilde{T}\Big( e^{i \int_0^t H_\chi[J^-(t')]dt'}\Big) ~~;~~H_\chi[J^\pm(t)] = H_{0 \chi}+\int d^3x J[\phi^\pm(t)]\mathcal{O}(\chi) \label{timevchi}\ee and $\widetilde{T}$ is the \emph{anti-time evolution operator} as befits   evolution backward in time. The calculation of the influence action is facilitated by passing to the interaction picture for the Hamiltonian $H_\chi[J(t)]$, defining
\be  \mathcal{U}(t;J^\pm) = e^{-i H_{0\chi}\,t} ~ \mathcal{U}_{ip}(t;J^\pm) \label{ipicture} \ee and the $e^{\pm i H_{0\chi}\,t}$ cancel out in the trace in (\ref{trasa}). Now the trace can be obtained systematically in perturbation theory in $g$. Up to $\mathcal{O}(g^2)$  we find
\bea \mathcal{F}[J^+,J^-] & = &    - g \int d^4x \Big( J^+[x]-J^-[x]\Big)\,\langle \mathcal{O}(x)\rangle \nonumber + \\ & & \frac{i g^2 }{2} \int d^4x_1 \int d^4x_2 \Bigg\{ J^+[x_1]\,J^+[x_2]\,G_c^{++}(x_1-x_2)+ J^-[x_1]\,J^-[x_2]\,G_c^{--}(x_1-x_2) \nonumber \\
 & - & J^+[x_1]\,J^-[x_2]\,G_c^{+-}(x_1-x_2)- J^-[x_1]\,J^+[x_2]\,G_c^{-+}(x_1-x_2)\Bigg\}\,. \label{finF}\eea In this expression $J^\pm[x]\equiv J[\phi^\pm(x)]$, and the \emph{connected} correlation functions are given by
\begin{eqnarray}
&& G_c^{-+}(x_1-x_2) =   \langle
{\cal O}(x_1) {\cal O}(x_2)\rangle - \langle \mathcal{O}(x_1)\rangle  \langle \mathcal{O}(x_2)\rangle =   {G}_c^>(x_1-x_2) \,,\label{ggreat} \\&&  G_c^{+-}(x_1-x_2) =   \langle
{\cal O}(x_2) {\cal O}(x_1)\rangle - \langle \mathcal{O}(x_2)\rangle  \langle \mathcal{O}(x_1)\rangle =   {G}_c^<(x_1-x_2)\,,\label{lesser} \\&& G_c^{++}(x_1-x_2)
  =
{ G}_c^>(x_1-x_2)\Theta(t_1-t_2)+ {  G}_c^<(x_1-x_2)\Theta(t_2-t_1) \,,\label{timeordered} \\&& G_c^{--}(x_1-x_2)
  =
{ G}_c^>(x_1-x_2)\Theta(t_2-t_1)+ {  G}_c^<(x_1-x_2)\Theta(t_1-t_2)\,,\label{antitimeordered}
\end{eqnarray} in terms of interaction picture fields, where
\be \langle (\cdots) \rangle = \mathrm{Tr}(\cdots)\rho_\chi(0)\,. \label{expec}\ee
Furthermore, for the case of hermitian operators $\mathcal{O}$ as considered here it follows that
\be G_c^>(x_1-x_2) = G_c^<(x_2-x_1)\,. \label{iden}\ee

The \emph{effective action} out of equilibrium is given by
\be  {S}_{eff}[\phi^+,\phi^-] = \int^t_0 dt \int d^3 x \Bigg\{ \mathcal{L}_0[\phi^+]-\mathcal{L}_0[\phi^-]\Bigg\} +\mathcal{F}[J[\phi^+],J[\phi^-]] \,.\label{Leff} \ee
Since   $\rho_\chi(0)=e^{-\beta H_{0\chi}}$ the expectation value $\langle \mathcal{O}(x_1)\rangle$ is independent of space-time coordinates. Therefore the first term on the right hand side of (\ref{finF}) can be cancelled by normal ordering the composite operator $\mathcal{O}(\chi)$ in the initial density matrix of the $\chi$ field, namely
\be \mathcal{O}(\chi) \rightarrow \mathcal{O}(\chi) - \langle \mathcal{O}(\chi)\rangle \,.\label{NO}\ee    This normal ordered operator features vanishing expectation value in the initial density matrix.  This is tantamount to introducing a counterterm in the Lagrangian density that cancels   the first term on the right hand side of (\ref{finF}) and the terms $\langle \mathcal{O}(x_{1,2})\rangle$ in the connected correlation functions (\ref{ggreat}-\ref{antitimeordered}), hence in what follows we suppress the subscript ``c'' in the correlation functions.

With the purpose of comparing the influence functional approach to the quantum master equation developed in section (\ref{sec:master}) and to exploit the spectral representation of the correlation functions, we re-write the influence action (\ref{finF}) solely in terms of the two correlation functions $G^\lessgtr$. This is achieved by implementing the following steps:

\begin{itemize}
\item{In the term with $J^+(x_1)J^+(x_2)$: in the contribution $G^<(x_1-x_2)\Theta(t_2-t_1)$ (see eqn. (\ref{timeordered})) relabel $t_1 \leftrightarrow t_2$ and use the property (\ref{iden}).  }
\item{ In the term with $J^-(x_1)J^-(x_2)$: in the contribution $G^>(x_1-x_2)\Theta(t_2-t_1)$ (see eqn. (\ref{antitimeordered})) relabel $t_1 \leftrightarrow t_2$ and use the property (\ref{iden}). }
    \item{ In the term with $J^+(x_1)J^-(x_2)$: multiply $G^<(x_1-x_2)$ by $\Theta(t_1-t_2)+\Theta(t_2-t_1)=1$ and in the term with $\Theta(t_2-t_1)$ relabel $ t_1 \leftrightarrow t_2$ and use the property (\ref{iden}). }
 \item{ In the term with $J^-(x_1)J^+(x_2)$: multiply $G^>(x_1-x_2)$ by $\Theta(t_1-t_2)+\Theta(t_2-t_1)=1$ and in the term with $\Theta(t_2-t_1)$ relabel $t_1 \leftrightarrow t_2$ and use the property (\ref{iden}). }
\end{itemize}
We find
\bea \mathcal{F}[J^+, J^-]  &  = &  i\,g^2\int d^3x_1 d^3x_2 \int^t_0 dt_1\,\int^{t_1}_0 dt_2\,\Bigg\{ J^+(\vx_1,t_1)J^+(\vx_2,t_2)\,G^>(x_1-x_2) +   J^-(\vx_1,t_1)J^-(\vx_2,t_2)\,G^<(x_1-x_2) \nonumber\\
  &- & J^+(\vx_1,t_1)J^-(\vx_2,t_2)\,G^<(x_1-x_2)  -   J^-(\vx_1,t_1)J^+(\vx_2,t_2)\,G^>(x_1-x_2)\Bigg\} \label{Funravel}\eea where $G^{\lessgtr}$ are given by eqns. (\ref{ggreat},\ref{lesser}).  This is the general form of the influence function up to second order in the system-environment coupling.

We note that
\bea \frac{d}{dt}\Bigg\{e^{i\mathcal{F}[J^+, J^-]}\Bigg\} & = &  -g^2\int d^3x_1 d^3x_2  \,\int^{t}_0 dt_2\,\Bigg\{ J^+(\vx_1,t)J^+(\vx_2,t_2)\,G^>(\vx_1-\vx_2,t-t_2) + \nonumber \\ && J^-(\vx_1,t)J^-(\vx_2,t_2)\,G^<(\vx_1-\vx_2,t-t_2)- J^+(\vx_1,t)J^-(\vx_2,t_2)\,G^<(\vx_1-\vx_2,t-t_2)\nonumber \\ & - & J^-(\vx_1,t)J^+(\vx_2,t_2)\,G^>(\vx_1-\vx_2,t-t_2)\Bigg\} ~\Bigg\{e^{i\mathcal{F}[J^+, J^-]}\Bigg\}\,, \label{dtF}\eea this result will be of paramount importance when  comparing to the quantum master equation in section (\ref{sec:master}).

\subsection{Spectral densities of the environment:}
The dynamics and dissipative processes depend on the correlation functions of the environment and crucially on their spectral density. In   appendix (A) we show that the correlation functions (\ref{ggreat}-\ref{lesser}) (and consequently (\ref{timeordered},\ref{antitimeordered}))  can be written in terms of a spectral representation, namely
\be G^\lessgtr(x-x') = \int \frac{d^4 k}{(2\pi)^4}\,\rho^\lessgtr(k_0,k)\,e^{-ik_0(t-t')}\,e^{i\vk\cdot(\vx-\vx')} \label{specrep}\ee where
\be \rho^>(k_0,k) = \rho(k_0,k)\big[1+n(k_0)] ~~;~~ \rho^<(k_0,k) = \rho(k_0,k) n(k_0) ~~;~~ n(k_0) = \frac{1}{e^{\beta k_0}-1} ~;~ \beta = 1/T \label{rhogl}\ee where $\rho(k_0,k)$ is the spectral density and $T$ is the temperature of the bath described by the $\chi$ field(s).  We analyze the different cases: \textbf{(a):}  one single $\chi$ field of mass $M_\chi \gg m_\phi$,   and $\mathcal{O}(\chi) = \chi$,  and \textbf{(b):} two different fields $\chi_1,\chi_2$ with $\mathcal{O}(\chi) = \chi_1\,\chi_2$ and masses $M_1> M_2 \gg m_\phi$.

The spectral densities for the different cases are obtained in detail in appendix (\ref{sec:specdens}), here we summarize their main features for the different cases.

\vspace{2mm}

\textbf{Case a): One $\chi$ field, $\mathcal{O}(\chi) = \chi$}

\vspace{2mm}

In this case it is straightforward to find (see Appendix B)
\be \rho(k_0,k) = \frac{\pi}{w_k}\,\Big[\delta(k_0-w_k)-\delta(k_0+w_k) \Big]~~;~~w_k= \sqrt{k^2+M^2}\,, \label{rho1chi}\ee  where $M$ is the mass of the $\chi$ field.

\vspace{2mm}

\textbf{Case b): Two $\chi$ fields, $\mathcal{O}(\chi) = \chi_1\,\chi_2$}

\vspace{2mm}

In Appendix B we provide the derivation of the spectral density at finite temperature $T$ for this case and show that it is of the form
\bea \rho(k_0,k;T) & = &  \rho_{LD}(k_0,k;T)_{LD}\,\Theta(-Q^2)+ \rho_D(k_0,k;T)\,\Theta((M_1-M_2)^2-Q^2)\,\Theta(Q^2)\nonumber \\ & + & \rho_{2\chi}(k_0,k;T)\,\Theta(Q^2-(M_1+M_2)^2) ~~;~~Q^2=k^2_0 - k^2 \label{rho2chis}\eea The expressions for the different $\rho's$ are given explicitly in appendix B and merit discussion.
\begin{itemize}
\item {The contribution $\rho_{LD}(k_0,k;T)$ with support below the light cone ($Q^2 <0$) corresponds to the process of Landau damping, this is a medium dependent contribution that vanishes in the $T\rightarrow 0$ limit. It describes collisionless damping in a medium as a consequence of dephasing.   While this contribution vanishes on the mass shell of the light field $\phi$ at $Q^2 =m^2_\phi$, it does contribute in the long time, long wavelength limit, as discussed in reference\cite{boylan} it is responsible for power law long time tails.}

 \item  {The contribution $\rho_D(k_0,k;T)$ also describes a process solely available in the medium corresponding to the \emph{decay}   $\chi_1 \rightarrow \chi_2 \phi$ (since $M_1 > M_2 \gg m_\phi$), this interpretation will be discussed in detail below. This part of the spectral density also vanishes for $T\rightarrow 0$, however at $T\neq 0$ it \emph{has support on the mass shell of the light field $\phi$}. This feature is important, as it will be discussed in detail below, it is responsible for the dissipative processes in the effective action and leads to \emph{thermalization} of the light field $\phi$ with the bath of heavy fields.  It may be argued that an effective field theory description would only be valid for $T\ll M_1, M_2$ so that the heavy fields are not excited in a thermal plasma, therefore these contributions are thermally suppressed in this regime. While this is correct, we will see below that even when the heavy degrees of freedom are thermally suppressed, \emph{the long time limit} will reveal full thermalization of the light field and if $m_\phi \ll T \ll M_1,M_2$ the light field will reach a large population as a consequence of the decay of the heavy degrees of freedom, even when these degrees of freedom are thermally suppressed. The thermal suppression implies longer time scales for relaxation towards thermal equilibrium and the population of the $\phi$ particles grows during this time scale from the decay of the heavy field.}

 \item {  The contribution $\rho_{2\chi}(k_0,k;T)$ corresponds to the usual two particle cut for $Q^2 > (M_1+M_2)^2$ and is finite in the $T\rightarrow 0$ limit where it is given by (see appendix B)
\be \rho(k_0,k;T=0) = \frac{\mathrm{sign}(k_0) }{8\pi\,Q^2}\Bigg\{\Big[Q^2-(M_1-M_2)^2 \Big]\,\Big[Q^2-(M_1+M_2)^2 \Big] \Bigg\}^{\frac{1}{2}}\,\Theta\Big[Q^2-(M_1+M_2)^2\Big]\, ~~;~~Q^2 = k^2_0-k^2  \label{zeroTrho2chi}\ee and $n(k_0) \rightarrow -\Theta(-k_0)$. The $\Theta$ function in (\ref{zeroTrho2chi}) corresponds to the two particle threshold. This is the \emph{only} contribution that survives at $T=0$ and leads to a local and unitary effective action in this limit. The threshold contribution to the noise kernel has been discussed previously only at $T=0$ and for $M_1=M_2$\cite{calhuthreshold}. }

\end{itemize}

\section{$\textbf{T=0}$: Long time limit, local effective field theory, thresholds and renormalization.}\label{sec:localqft}

Considering the factorized initial density matrix (\ref{inidensmtx}) and evolving it in time with the full interacting Hamiltonian (\ref{unitimeop},\ref{hami})  is tantamount to a sudden ``switch-on'' of the interaction  resulting in transient non-equilibrium dynamics which is described by the effective action (\ref{Leff}). While the initial transients arising from the sudden ``switching-on'' of the interaction are of timely interest in a wide range of experimentally available phenomena in condensed matter physics, in this section we focus on the long-time behavior after the initial transient phenomena has relaxed. Our goal here is to study under what circumstances the long time limit after the transient dynamics has relaxed the time evolution of the reduced density matrix is described by an effective \emph{local} Hamiltonian dynamics yielding \emph{unitary} time evolution. We first consider $T=0$, in this case the ``bath'' or environment is in its \emph{ground state} and the spectral density of correlation functions of the bath degrees of freedom  features \emph{only} the usual multiparticle thresholds.  From this study we draw conclusions on dissipative aspects which will highlight the influence of the medium at $T\neq 0$.

In order to relate the long time behavior of the influence functional to the spectral representation of the bath correlation functions we define
\be J^\pm(\vk,t) = \int d^3 x e^{i\vk\cdot\vx} \,J^\pm(\vx,t) \label{sft}\ee keeping the same notation for the spatial Fourier transform, secondly, for the $J^\pm$ with time argument $t_2$ we
introduce (for the spatial Fourier transforms)
\be J^\pm(\vk,t) = \int \frac{dp_0}{2\pi}\,e^{-ip_0 t}\, J^\pm(\vk,p_0)\,,  \label{jsofpo}\ee finally we carry out the integral over $t_2$
\be \int^{t_1}_0 e^{i(p_0-k_0)(t_1-t_2)}\,dt_2 = i~ \frac{\Big[1-\cos[(p_0-k_0)t_1] \Big]}{(p_0-k_0)}+  \frac{\sin[(p_0-k_0)t_1]}{(p_0-k_0)} ~~~ \overrightarrow{t_1 \rightarrow \infty}~~~i \mathcal{P}\Big[ \frac{1}{p_0-k_0} \Big]+ \pi \delta(p_0-k_0)\label{t2int}\ee
where $\mathcal{P}$ stands for the principal part and the equivalence on the right hand side is understood when integrated with a smooth density of states\footnote{Alternatively take $t_1\rightarrow \infty$ with a convergence factor $\epsilon \rightarrow 0^+$ to yield $\frac{i}{p_0-k_0+i\epsilon}$.}. The influence function is given by
\be \mathcal{F}[J^+,J^-] = \mathcal{F}_H[J^+,J^-] + \mathcal{F}_D[J^+,J^-]\,, \label{FHD}\ee
  introducing
\be \Delta^\lessgtr (p_0,k) =     \int  \frac{dk_0}{2\pi} ~\mathcal{P}\Bigg[ \frac{\rho^\lessgtr (k_0,k)}{p_0-k_0}\Bigg] \label{Deltagl}\ee we find
\bea   \mathcal{F}_H[J^+,J^-]   & = &   - g^2 \int \frac{d^3k}{(2\pi)^3}\int \frac{dp_0}{(2\pi)}\int_0^{t}dt_1\, e^{-ip_0 t_1}\Bigg\{ J^+(\vk,t_1)J^+(-\vk,p_0)\Delta^>(p_0,k) + \nonumber \\ &&  J^-(\vk,t_1)J^-(-\vk,p_0)\Delta^<(p_0,k) - J^+(\vk,t_1)J^-(-\vk,p_0)\Delta^<(p_0,k)-\nonumber \\ & & J^-(\vk,t_1)J^+(-\vk,p_0)\Delta^>(p_0,k) \Bigg\}\,, \label{Fham}\eea
 and
\bea   \mathcal{F}_D[J^+,J^-]   & = &   i\, {g^2}{\pi}  \int \frac{d^3k}{(2\pi)^3}\int \frac{{dp_0}}{2\pi} \int_0^{t}dt_1\, e^{-ip_0 t_1}\Bigg\{ J^+(\vk,t_1)J^+(-\vk,p_0)\rho^>(p_0,k) + \nonumber \\ &&  J^-(\vk,t_1)J^-(-\vk,p_0)\rho^<(p_0,k) - J^+(\vk,t_1)J^-(-\vk,p_0)\rho^<(p_0,k)-\nonumber \\ & & J^-(\vk,t_1)J^+(-\vk,p_0)\rho^>(p_0,k) \Bigg\}\,. \label{Fdis}\eea
These contributions describe very different processes: $ \mathcal{F}_H[J^+,J^-]$ describes the contribution to the effective action from virtual intermediate states of the ``bath'' fields, as shown below, it features a non-vanishing local limit  when $p_0,k$  are well below the threshold for excitation of the ``environmental'' degrees of freedom. This contribution describes the ``fluctuating'' part of the effective action. Instead $\mathcal{F}_D[J^+,J^-] $ describes dissipative processes. For this contribution to be non-vanishing,   $p_0,k$ must be within the region of support of the spectral density, namely above the multiparticle thresholds. This contribution to the effective action describes production or decay of the environmental degrees of freedom namely the transfer of energy and momentum between the system degrees of freedom to those of the bath, resulting in a dissipative contribution to the effective action.

 This interpretation becomes clear by considering the two cases (\ref{rho1chi},\ref{zeroTrho2chi}) for the spectral density at $T=0$.

\vspace{2mm}

\textbf{Case a)  $ g J[\phi]\,\chi$: } in this case for $T=0$ it follows from (\ref{rhogl},\ref{rho1chi}) that
\be \rho^>(k_0,k)= \frac{\pi}{w_k}\,\delta(k_0-w_k)~;~\rho^<(k_0,k)= \frac{\pi}{w_k}\,\delta(k_0+w_k) \,,\label{rhosis} \ee therefore
\be   \Delta^>(p_0,k) = \frac{1}{2w_k} \mathcal{P}\Big[\frac{1}{p_0-w_k} \Big]~~;~~ \Delta^<(p_0,k) = \frac{1}{2w_k} \mathcal{P}\Big[\frac{1}{p_0+w_k} \Big] \,, \label{intsdk0}\ee

If the currents $J^{\pm}(\vk,p_0)$ \emph{only have support for $k,p_0 \ll M$ } ($M$ is the mass of the single $\chi$ field) then (\ref{intsdk0})  can be replaced by $-1/(2M^2)$ and $ 1/(2M^2)$ respectively and taken out of the integrals in (\ref{Fham}). Using (\ref{sft}) and (\ref{jsofpo}) we find that the last two terms (\ref{Fham}) cancel out, leading to
\be \mathcal{F}_{H}[J^+,J^-] = \frac{g^2}{2M^2} \int d^3x \int^t_0 dt_1\,\Bigg\{ \big(J^+(\vx,t_1)\big)^2 -  \big(J^-(\vx,t_1)\big)^2 \Bigg\}\,, \label{fhamlocal}\ee furthermore, for $p_0 \ll M  $ it follows that $\rho^\lessgtr (p_0, k) =0$ since the spectral density does not have support for $p_0 \neq \pm w_k$, therefore
\be \mathcal{F}_D[J^+,J^-] =0\,. \label{Fdiszero}\ee

In this case the total effective action (\ref{Leff}) becomes
\be  {S}_{eff}[\phi^+,\phi^-] = \int^t_0 dt_1 \int d^3 x \Bigg\{ \Big[\mathcal{L}_0[\phi^+]+ \frac{g^2}{2M^2}(J[\phi^+])^2\Big] -\Big[\mathcal{L}_0[\phi^-]+ \frac{g^2}{2M^2}(J[\phi^-])^2\Big]\Bigg\}\,.\label{LeffFham} \ee This effective action describes a local  effective field theory leading to \emph{unitary} time evolution of the density matrix of the field $\phi$ with the effective local Hamiltonian
\be H_{eff} = H_{0\phi}-\frac{g^2}{2M^2}\int d^3 x (J[\phi])^2\,. \label{heffunitary}\ee
 For example, with $J[\phi]=\phi^2$ this result simply describes an effective local ``Fermi theory'' by integrating out the heavy degree of freedom in the intermediate state, depicted in fig. (\ref{fig:effvera}). The result is expected from carrying out the Gaussian path integrals over $\chi$ \emph{without} the kinetic and spatial gradient terms as befits nearly zero frequency and momentum transfer. It is reassuring that it emerges in the local limit of the non-equilibrium effective action.

\begin{figure}[h!]
\includegraphics[height=3.5in,width=3.5in,keepaspectratio=true]{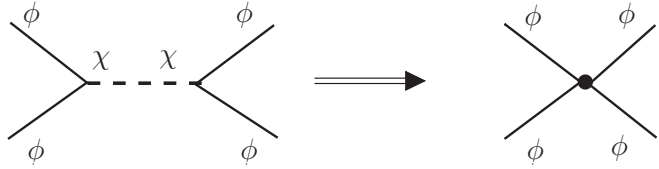}
\caption{Effective vertex in case \textbf{a)} for $J[\phi]=\phi^2$.}
\label{fig:effvera}
\end{figure}

The term $\mathcal{F}_D[J^+,J^-]$ given by (\ref{Fdis}) leads to non-unitary time evolution and is purely dissipative, it describes the coupling of the field $\phi$ to the ``continuum'' of the heavy field(s) as gleaned from the fact that the spectral density is evaluated at the ``external'' frequency $p_0$ at momentum $k$. In this simple case the ``continuum'' is just one field mode and the spectral density vanishes for $p_0,k \ll M$, there is no dissipation in this case and the reduced density matrix evolves in time with \emph{unitary} time evolution, namely
\be \rho^r(t) = e^{-iH_{eff}t}~\rho^r(0) ~ e^{iH_{eff}t}\,. \label{unitaror}\ee

\vspace{2mm}

\textbf{Case b):} for  this case when $T=0$ it follows from (\ref{rhogl},\ref{zeroTrho2chi}) that
\be \rho^>(k_0,k;T=0) =  \Theta(k_0)\,\big|\rho(k_0,k;T=0)\big|~;~
\rho^<(k_0,k;T=0) =  \Theta(-k_0)\,\big|\rho(k_0,k;T=0)\big|\,. \label{rhogl2chis} \ee This spectral density describes a two particle continuum above the thresholds $|k_0| \geq q_{th}=\sqrt{k^2+(M_1+M_2)^2}$. Therefore we find
\be \Delta^>(p_0;k) = \int_{q_{th}}^\infty dk_0 ~ \mathcal{P}\Bigg[\frac{\big|\rho(k_0,k;T=0)\big|}{p_0-k_0} \Bigg]~~;~~\Delta^<(p_0;k) = \int_{q_{th}}^\infty dk_0 ~ \mathcal{P}\Bigg[\frac{\big|\rho(k_0,k;T=0)\big|}{p_0+k_0} \Bigg]\,. \label{deltaths} \ee

We note that as $k_0 \rightarrow \infty $ it follows from (\ref{zeroTrho2chi}) that $\big|\rho(k_0,k;T=0)\big| \rightarrow 1/8\pi$ therefore $\Delta^\lessgtr (p_0,k)$ become  logarithmically divergent in the ultraviolet. This is precisely the zero temperature renormalization of the effective ``current-current'' vertex from the two particle intermediate state.

 As we are considering a ``low energy'' effective quantum field theory of a light field after integrating out the heavy fields, all the correlation functions of the light field are restricted to feature transferred momenta well below the two particle continuum, in particular $|p_0|, q \ll q_{th} $ therefore $\rho^\lessgtr (p_0,k)=0$, and we can take $p_0,k \rightarrow 0$  in (\ref{deltaths}), to extract the local limit. Because $\Delta^>(0;0) = -\Delta^<(0;0)$ we find again that the last two terms in (\ref{Fham}) cancel out and   using (\ref{sft}, \ref{jsofpo}) we find
\bea \mathcal{F}_{H}[J^+,J^-] & = &  g^2\,Z_g \int d^3x \int^t_0 dt_1\,\Bigg\{ \big(J^+(\vx,t_1)\big)^2 -  \big(J^-(\vx,t_1)\big)^2 \Bigg\} \nonumber \\\mathcal{F}_{D}[J^+,J^-]&  = &  0\,, \label{fhamlocalcaseb}\eea where
\be Z_g =  \int_{q_{th}}^\infty \frac{dk_0}{k_0} ~   \big|\rho(k_0,0;T=0)\big|  \label{reng}\ee is ultraviolet logarithmically divergent. For $J[\phi] = \phi^2$ the local effective vertex is shown in fig.(\ref{fig:effverb}), and the ultraviolet logarithmic divergence corresponds to coupling renormalization.

For the case $J[\phi]=\phi$ the local effective vertex is shown in fig.(\ref{fig:mass}), the loop corresponds to a self energy correction and includes mass and wavefunction  renormalizations. To lowest order in $p_0,k$, the effective action becomes
\be  {S}_{eff}[\phi^+,\phi^-] = \int^t_0 dt_1 \int d^3 x \Bigg\{ \Big[\mathcal{L}_0[\phi^+]+ g^2\,Z_g~(J[\phi^+])^2\Big] -\Big[\mathcal{L}_0[\phi^-]+ g^2\,Z_g~(J[\phi^-])^2\Big]\Bigg\}\,.\label{LeffFhamb} \ee It describes  unitary time evolution as (\ref{unitaror})  in the asymptotically long time limit after the transient dynamics has relaxed.

\begin{figure}[h!]
\includegraphics[height=3.5in,width=3.5in,keepaspectratio=true]{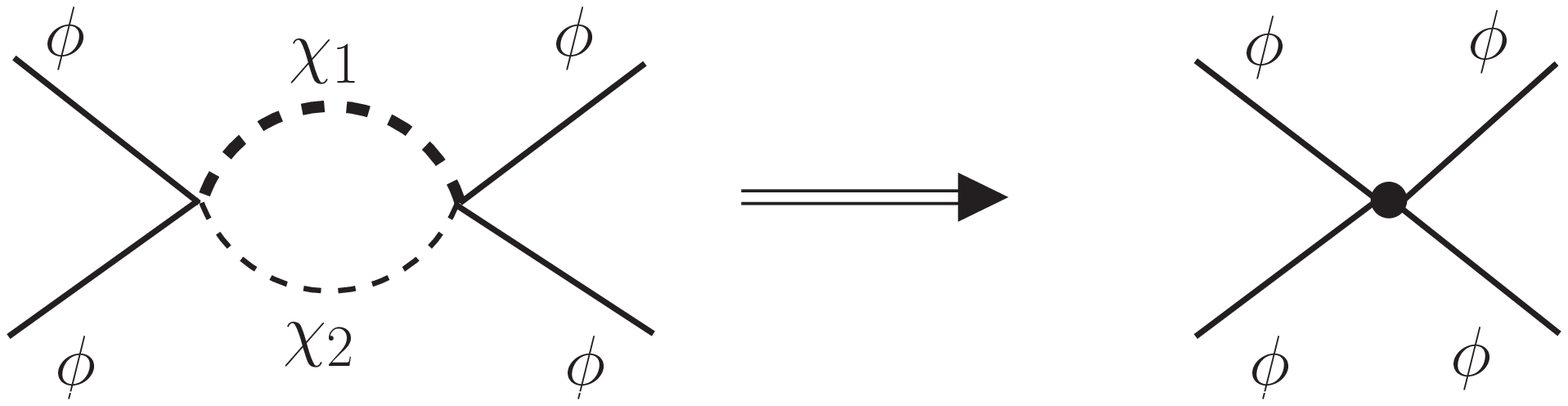}
\caption{Effective vertex in case \textbf{b)} for $J[\phi]=\phi^2$.}
\label{fig:effverb}
\end{figure}

\begin{figure}[h!]
\includegraphics[height=3.5in,width=3.5in,keepaspectratio=true]{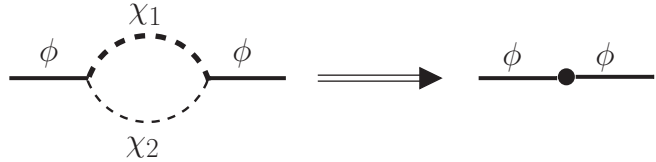}
\caption{Effective vertex in case \textbf{b)} for $J[\phi]=\phi^2$.}
\label{fig:mass}
\end{figure}

Although we focused on the limits $p_0, k \simeq 0$ to extract the local limit, for $p_0 \ll q_{th}$ we can go beyond and expand in $p_0, k$ thereby generating a derivative expansion of the effective action  in powers of $\big(\partial_\mu\,\phi/(M_1+M_2)\big)^2$ that describes the asymptotic long-wavelength and low frequency dynamics with a local effective   action.

In the case of $J[\phi]=\phi$ this will also contain the (finite) wave function renormalization. This is the   dynamics expected from a local effective field theory, it is manifestly unitary because the dissipative term vanishes identically.

This local effective and \emph{unitary} description is valid for long wavelength and low frequency phenomena so that $p_0,k$ are well below the threshold   to excite the environmental degrees of freedom. The description in terms of a local effective field theory emerges in the asymptotic long time limit well after transient phenomena has relaxed.

 When $p_0$ becomes larger than multiparticle threshold the dissipative contribution to the influence action, $\mathcal{F}_D\neq 0$   as a consequence of particle production and the time evolution of the reduced density matrix is no longer unitary.

For $T\neq 0$ new contributions to the two particle spectral density arise as shown explicitly in eqn. (\ref{rho2chisapp},\ref{rhold},\ref{rhode})   in appendix (\ref{sec:specdens}). These have support \emph{below} the $T=0$ two particle threshold and yield new phenomena. In particular, if $(M_1-M_2)^2 > (p_0^2-k^2)$ the contribution from $\rho_D$ (which vanishes for $T=0$) in  the spectral density (\ref{rho2chis}) (see \ref{rho2chisapp}-\ref{rho2chitos} for the explicit contributions) will lead to dissipative processes and non-unitary time evolution which is studied below.

\section{ Stochastic description: Brownian fields and Langevin equation. }\label{sec:langevin}

In this section we study new dynamical phenomena associated with the $T\neq0$ contributions to the spectral density with new multiparticle cuts below the two particle continuum.

In order to establish a clear relation to a stochastic description  we study the case
\be J[\phi(x)] = \phi(x)~~;~~\mathcal{O}(\chi) = \chi_1\,\chi_2 \,,\label{linearJ}\ee which is   replaced in the general effective functional
  eqn. (\ref{finF}) (  $\langle \mathcal{O}(x)\rangle =0$).

 Correlation functions of the fields $\phi^\pm$ along the forward and backward branches can be obtained by introducing sources $h^\pm(\vx, t) $ linearly coupled to these fields and taking functional derivatives with respects to these sources. Therefore the effective action (\ref{Leff}) is generalized to
\be  {S}_{eff}[\phi^\pm;h^\pm] = \int^\infty_0 dt \int d^3 x \Bigg\{ \mathcal{L}_0[\phi^+]+h^+(\vx,t)\phi^+(\vx,t)-\mathcal{L}_0[\phi^-]-h^-(\vx,t)\phi^-(\vx,t)\Bigg\} +\mathcal{F}[J[\phi^+],J[\phi^-]]\,. \label{Leffhs} \ee We have taken the upper time limit $t\rightarrow \infty$ because we can obtain the correlation functions of the fields $\phi^\pm$ at arbitrary times by taking functional derivatives with respect to the sources $h^\pm$ at different times.

Therefore we consider the generating functional
\be \mathcal{Z}[h^\pm] = \int D\phi_f ~\rho^r(\phi_f,\phi_f,\infty,h^\pm) \,,\label{generah}\ee
where $\rho^r$ is given by eqn. (\ref{rhored}) with the addition of the sources $h^\pm$ in the effective action.

We introduce the   center of mass and
relative variables \be \Psi(\vec x,t)   =   \frac{1}{2}
\left(\phi^+(\vec x,t) + \phi^-(\vec x,t) \right) \; \; ; \; \;
R(\vec x,t) = \left(\phi^+(\vec x,t) - \phi^-(\vec x,t) \right)
\label{wigvars} \ee and similarly for the sources
\be H = \frac{1}{2}(h^++h^-)~~;~~h = (h^+-h^-) \label{sources}\ee

\noindent  and the Wigner transform of the initial density matrix $\hat{\rho}_\phi(0)$
of the $\phi$ field

\begin{equation}
{\cal W}(\Psi_i ; \Pi_i) = \int D R_i e^{-i\int d^3x \Pi_i(\vec x)
R_i(\vec x)} \rho_{\phi,0}(\Psi_i+\frac{R_i}{2};\Psi_i-\frac{R_i}{2})\,. \label{wignerrho}\ee

The Wigner transform leads to a quasiclassical description of the dynamics, it is the closest ``proxy'' to a (semi) classical phase space distribution\cite{schmid,calhubuk}, its time evolution is the Fokker-Planck equation\cite{calhubuk}. Accordingly, the center of mass combination $\Psi$ is the closest to a semiclassical description of the field\cite{schmid}, in fact
\be \mathrm{Tr} \phi ~ \hat{\rho}^r = \mathrm{Tr} \Psi ~ \hat{\rho}^r ~~;~~ \mathrm{Tr} R ~ \hat{\rho}^r  =0 \,. \label{avscm}\ee

The inverse transform of the Wigner function is given by
\be \rho_{\phi,0}(\Psi_i+\frac{R_i}{2};\Psi_i-\frac{R_i}{2})= \int D \Pi_i
e^{i\int d^3x \Pi_i(\vec x) R_i(\vec x)}{\cal W}(\Psi_i ;
\Pi_i)\,.\label{inverwignerrho}
\end{equation}
The boundary conditions on the $\phi^\pm$ path integrals given by
(\ref{piforwardbc},\ref{pibackbc}) translate into the following boundary conditions on
the center of mass and relative variables
\begin{equation}
\Psi(\vec x,t=0)= \Psi_i \; \; ; \; \; R(\vec x,t=0)= R_i
\label{bcwig}
\end{equation}

\noindent furthermore, the trace in $\mathcal{Z}$ entails that $\phi_f=\phi'_f$ leading to the constraint
\be   R(\vec{x},t=\infty)= 0 \label{Rinfty} \,. \ee

 In
terms of the spatial Fourier transforms of the center of mass and
relative  variables  introduced above  (\ref{wigvars}), integrating by
parts and accounting for the boundary conditions (\ref{bcwig}), the
non-equilibrium effective action (\ref{Leffhs}) with $J[\phi]=\phi$ in eqn. (\ref{finF}) becomes:
\begin{eqnarray}
iS_{eff}[\Psi,R] & = & \int_0^{\infty} dt \sum_{\vec k} \left\{-i R_{-\vec k}\left( \ddot{\Psi}_{\vec k}(t)+\Ok^2\,\Psi_{\vk}(t)-H_{\vk}(t) \right)+ \Psi_{\vk}h_{-\vk}\right\} \nonumber \\
                 & - & \int_0^{\infty} dt \int_0^{\infty} dt' \left\{\frac{1}{2}\,R_{\vec k}(t')\,{\mathcal{N} }_k(t-t')\,R_{-\vec k}(t) + R_{-\vec k}(t)\,
i\Sigma^R_k(t-t')\, \Psi_{\vec k}(t') \right \} \nonumber \\
& + & \int d^3x R_i(\vec x) \dot{\Psi}(\vec x,t=0)
\label{efflanwig}
\end{eqnarray}
where $\Ok^2=k^2+m^2_\phi$, and the last term arises after  integration by parts in time,
using the boundary conditions (\ref{bcwig}) and (\ref{Rinfty}).
The kernels in the above effective Lagrangian are given by (see
eqns. (\ref{ggreat}-\ref{antitimeordered}))
\begin{eqnarray}
\mathcal{N}_k(t-t') & = &
\frac{g^2}{2} \left[   G^>(k;t-t')+{ G}^<(k;t-t') \right] \label{kernelkappa} \\
i\Sigma^{R}_k(t-t') & = &  g^2 \left[{ G}^>(k;t-t')-{
G}^<(k;t-t') \right]\Theta(t-t') \equiv
i\Sigma_{k}(t-t')\Theta(t-t') \label{kernelsigma}
\end{eqnarray} where $G^{<,>}(k;t-t')$ are the spatial Fourier transforms of the correlation functions in (\ref{ggreat}-\ref{antitimeordered}).

The term quadratic in the relative variable $R$ can be written in terms of a stochastic
noise as
\begin{eqnarray}
\exp\Big\{-\frac{1}{2} \int dt \int dt' R_{-\vec k}(t)\mathcal{N}_k(t-t')R_{\vec k}(t')\Big\} & = & \int {\cal D}\xi
\exp\Big\{-\frac{1}{2} \int dt \int dt' ~~ \xi_{\vec k}(t)
\mathcal{N}^{-1}_k(t-t')\xi_{-\vec k}(t')  \nonumber \\
 &+ & i \int dt ~~ \xi_{-\vec k}(t) R_{\vec k}(t)\Big\}
\label{noisefunc}
\end{eqnarray}

We now set the sources $h^\pm =0 ~(H=h=0)$ to simplify the discussion, they can be included a posteriori to obtain the correlation functions.

The non-equilibrium generating functional can now be written in the
following form
\begin{eqnarray}
{\cal Z}  & = &   \int D \Psi_i \int D \Pi_i \int {\cal D} \Psi {\cal D}R {\cal D}\xi ~~
{\cal W}(\Psi_i;\Pi_i) DR_i
 e^{i \int d^3x R_i(\vec x) \left(\Pi_i (\vec x)-\Psi(\vec x,t=0)\right)}~
 {\cal P}[\xi]\, \times \nonumber \\
&  & \exp\left\{-i \int_0^{\infty} dt ~~ R_{-\vec k}(t) \left[
\ddot{\Psi}_{\vec k}(t)+\Ok^2\Psi_{\vec k}(t)+\int_0^t dt' ~~ \Sigma^{R}_k(t-t')\Psi_{\vec k}(t')-\xi_{\vec k}(t) \right] \right\} \eea where
\be{\cal P}[\xi]  =   \exp\left\{-\frac{1}{2} \int_0^{\infty} dt
\int_0^{\infty} dt' ~~ \xi_{\vec k}(t) \mathcal{N}^{-1}_k(t-t')
\xi_{-\vec k}(t') \right\} \,. \label{genefunc}
\ee

The functional integral over $R_i$ can now be done, resulting in a functional delta function,
that fixes the boundary condition $\dot{\Psi}(\vec x,t=0) =
\Pi_i(\vec x)$.

Finally the path integral over the relative variable can be
performed, leading to a functional delta function and the final form
of the generating functional  given by
\begin{eqnarray}
{\cal Z}   =   \int D \Psi_i   D \Pi_i ~~{\cal W}(\Psi_i;\Pi_i)
 {\cal D} \Psi {\cal D}\xi ~~ {\cal P}[\xi]\,
 \delta\left[
\ddot{\Psi}_{\vec k}(t)+\Ok^2\Psi_{\vec k}(t)+\int_0^{t} dt'
~\Sigma_k(t-t')\Psi_{\vec k}(t')-\xi_{\vec k}(t) \right]
\label{deltaprob}
\end{eqnarray}
with the boundary conditions on the path integral on $\Psi$ given by
\begin{equation}
\Psi(\vec x,t=0) = \Psi_i(\vec x) \; \; ; \; \; \dot{\Psi}(\vec
x,t=0)= \Pi_i(\vec x) \label{bcfin}
\end{equation}
\noindent where we have used the definition of $\Sigma^{R}_k(t-t')$
in terms of $\Sigma_k(t-t')$ given in equation (\ref{kernelsigma}).

The meaning of the above generating functional is the following: in
order to obtain correlation functions of the center of mass Wigner
variable $\Psi$ we must first find the solution of the classical
{\em stochastic} Langevin equation of motion
\begin{eqnarray}
&& \ddot{\Psi}_{\vec k}(t)+\Ok^2\Psi_{\vec k}(t)+\int_0^t dt'
~ \Sigma_{k}(t-t')
\Psi_{\vec k}(t')=\xi_{\vec k}(t) \nonumber \\
&& \Psi_{\vec k}(t=0)= \Psi_{i,\vec k} ; ~~
\dot{\Psi}_{\vec k}(t=0)= \Pi_{i,\vec k}
\label{langevin}
\end{eqnarray}
for arbitrary noise term $\xi$ and then average the products of
$\Psi$ over the stochastic noise with the Gaussian probability
distribution ${\cal P}[\xi]$ given by (\ref{genefunc}), and finally
average over the initial configurations $\Psi_i(\vec x);
\Pi_i(\vec x)$ weighted by the Wigner function ${\cal
W}(\Psi_i,\Pi_i)$, which plays the role of an initial (semiclassical)
 phase space distribution function.

   There are two different averages:

\begin{itemize}
\item{ The average over the  stochastic noise term, which up to
this order is Gaussian. We denote the average of a functional
$\mathcal{A}[\xi]$  over the noise with the probability distribution
function $P[\xi]$ given by eqn. (\ref{genefunc}) as

\begin{equation}\label{stocha}
\langle \langle \mathcal{A}[\xi] \rangle \rangle \equiv \frac{\int
\mathcal{D}\xi P[\xi] \mathcal{A}[\xi]}{\int \mathcal{D}\xi P[\xi]}.
\end{equation}

Since the noise probability distribution function is Gaussian the
only necessary correlation functions for the noise are given by

\begin{equation}
\langle \langle \xi_{\vec{k}}(t)\rangle \rangle =0 \; , \; \langle
\langle \xi_{\vec{k}}(t)\xi_{-\vec{k}'}(t')\rangle \rangle =
 \mathcal{N}_k(t-t')\,\delta^{3}(\vec{k}-\vec{k}')
\label{noisecorrel}
\end{equation}
\noindent and  the higher order correlation functions are obtained
from Wick's theorem. In general the  noise kernel
$\mathcal{N}_k(t-t')\neq \delta(t-t')$ namely
\emph{colored} noise.
From the results in appendix (\ref{app:fd}) we find
\bea i\Sigma_{k}(t-t')  & = &  g^2 ~ \int \frac{dk_0}{(2\pi)}\,\rho(k_0,k) \, e^{-ik_0(t-t')}  \label{isig} \\
\mathcal{N}_k(t-t') & = & \frac{g^2}{2}~  \int \frac{dk_0}{(2\pi)}\,\rho(k_0,k) \,\mathrm{coth\big[ \frac{\beta k_0}{2}\big]} e^{-ik_0(t-t')}  \label{noiset}\eea therefore the self energy $i\Sigma$ and noise $\mathcal{N}$ kernels obey the generalized fluctuation dissipation relation derived in appendix (\ref{app:fd}).

In the literature it is usually \emph{assumed} that the noise term $\mathcal{N}_k(t-t')$ has very short range time correlations, namely
\be \mathcal{N}_k(t-t') \propto \delta(t-t') \label{delnoise}\ee this assumption entails that
\be \rho(k_0,k) \,\mathrm{coth\big[ \frac{\beta k_0}{2}\big]} \simeq \mathrm{constant}\,,\label{const} \ee in the classical limit $\mathrm{coth[k_0/2T]}\simeq 2T/k_0$ this implies an ohmic spectral density, namely \be \rho(k_0,k) \propto k_0 \label{ohm} \ee however, this spectral density is not compatible with that arising in the relativistic field theory from the intermediate state with heavy particles not even as an approximation within a range of frequencies. In particular the presence of thresholds implies typically power law long time tails and at finite temperature the ``anomalous'' thresholds and the non-vanishing support of the spectral densities below the $T=0$ multiparticle thresholds clearly suggest that the ohmic spectral density cannot reliably describe the temporal correlations of a realistic bath.

While such an approximation may lead to an agreement with the long time dynamics in special cases, it is generally a crude approximation that merits scrutiny in each case.
 }

\item{The average over the initial conditions with the Wigner
distribution function ${\cal W}(\Psi_i,\Pi_i)$ which we denote as

\begin{equation}
\overline{\mathcal{A}[\Psi_i,\Pi_i]} \equiv  \frac{\int D \Psi_i
\int D\Pi_i ~~{\cal W}(\Psi_i;\Pi_i) \mathcal{A}[\Psi_i,\Pi_i]}{\int
D \Psi_i \int D\Pi_i ~~{\cal W}(\Psi_i;\Pi_i) } \label{iniaverage}
\end{equation}
For example for a Gaussian initial density matrix with vacuum free field correlations and anticipating a mass renormalization,  the Wigner
distribution function yields  the following averages:
\bea
&& \overline{\Psi_{i,\vec k}\Psi_{i,-\vec k}} =
\frac{1}{2\Okr} + \overline{\Psi_{i,\vk}}~~\overline{\Psi_{i,-\vk}}  \label{psi2}\\
&&\overline{\Pi_{i,\vec k}\Pi_{i,-\vec k}} =
\frac{\Okr}{2} + \overline{\Pi_{i,\vk}}~~\overline{\Pi_{i,-\vk}}\label{pi2} \\
&& \overline{\Pi_{i,\vec k}\Psi_{i,-\vec k}+\Psi_{i,\vec k}
\Pi_{i,-\vec k}} =  \overline{\Pi_{i,\vec k}}~~\overline{\Psi_{i,-\vec k}}+
\overline{\Psi_{i,\vec k}}~~\overline{\Pi_{i,-\vec k}}  \,, \label{psipi} \eea where $\Okr$ are the renormalized frequencies. Renormalization is discussed below.
}

\end{itemize}

The average in the time evolved full density matrix is therefore
defined by

\begin{equation}
  \overline{\langle \langle
\mathcal{A} \rangle \rangle}= \equiv \frac{\int
\mathcal{D}\xi P[\xi]~ \overline{\mathcal{A}[\Psi,\Pi,\xi]}}{\int \mathcal{D}\xi P[\xi]}\,. \label{totave}
\end{equation}

 Calling
the solution of (\ref{langevin}) $\Psi_{\vec
k}(t;\xi;\Psi_i;\Pi_i)$,  the two point correlation function of $\Psi$  is given by
\begin{equation}
 \overline{\langle \langle \Psi_{-\vec k}(t) \Psi_{\vec k}(t') \rangle \rangle}  \,. \label{expecvalcm}
\end{equation} It is straightforward to confirm that the equal time correlation function of the field $\phi$ is given by
\be \langle \phi_{\vk}(t)\phi_{-\vk}(t) \rangle =  \overline{\langle \langle \Psi_{-\vec k}(t) \Psi_{\vec k}(t) \rangle \rangle}  \,.  \label{eqtimfi}\ee where the average on the left hand side of this equation is in the reduced density matrix.

This result is  remarkably similar to the Martin-Siggia-Rose\cite{msr}   formulation for
stochastic \emph{classical} systems.

The solution of the Langevin equation (\ref{langevin}) is obtained by a Laplace transform as befits an initial value problem.   Defining the Laplace transforms

\begin{eqnarray}
\widetilde{\Psi}_{\vec k}(s)  \equiv \int^{\infty}_0 dt
e^{-st}\Psi_{\vec k}(t)\label{laplapsi}\\
\widetilde{\xi}_{\vec k}(s)  \equiv \int^{\infty}_0 dt
e^{-st}\xi_{\vec k}(t)\label{laplaxi}
\end{eqnarray}

\begin{equation}
\widetilde{\Sigma}(k,s)\equiv \int^{\infty}_0 dt e^{-st}\Sigma_k(t)=
-\frac{g^2}{2\pi} \int^{\infty}_{-\infty}
\frac{\rho(k_0,k)}{k_0-is}~dk_0\,.
\label{laplasig}
\end{equation}

\noindent in terms of which we find

\begin{equation}\label{solutionlap}
\widetilde{\Psi}_{\vec k}(s)=\frac{\Pi_{i,\vec k}+s\Psi_{i,\vec
k}+\widetilde{\xi}_{\vec
k}(s)}{s^2+\Ok^2+\widetilde{\Sigma}(k,s)}
\end{equation}
\noindent where we have used the initial conditions (\ref{bcfin}).
  Introducing the function ${G}_k(t)$ that obeys the
following equation of motion and initial conditions

\begin{equation}
\ddot{ G}_{k}(t)+\Ok^2\,{G}_{ k}(t)+\int_0^t dt' ~
\Sigma_{k}(t-t') G_{ k}(t')=0~~;~~  {G}_k(t=0)= 0; ~~ \dot{{G}}_{
k}(t=0)=1 \label{functionf}
\end{equation}

\noindent whose Laplace transform is given by

\begin{equation}\label{laplaf}
\widetilde{G}_k(s) =
\frac{1}{s^2+\Ok^2+\widetilde{\Sigma}(k,s)}
\end{equation}
we find that the solution of the Langevin
equation (\ref{langevin}) in real time is given by

\begin{equation}
\Psi_k(t;\Psi_i;\Pi_i;\xi) = \Psi_{i,\vec k}~ \dot{G}_k(t) +
 \Pi_{i,\vec k}~ G_k(t)+ \int^t_0
G_k(t-t')~\xi_{\vec k}(t') dt' \,,\label{inhosolution}
\end{equation} the last term highlights the stochastic nature of the effective field theory.

The real time solution for $G_k(t)$ is found by the inverse Laplace
transform

\begin{equation}\label{bromw}
G_k(t) = \int_{C}\frac{ds}{2\pi i} \frac{e^{st}
}{s^2+\Ok^2+\widetilde{\Sigma}(k,s)}
\end{equation}
\noindent where $C$ stands for the Bromwich contour, parallel to the
imaginary axis in the complex $s$ plane to the right of all the
singularities of $\widetilde{G}_k(s)$ and along the semicircle at
infinity for $\mathrm{Re}\,s < 0$. The singularities of
$\widetilde{G}_k(s)$ in the physical sheet are isolated single
particle poles  and multiparticle cuts along the imaginary axis. Decaying states correspond to complex poles with a \emph{negative} real part. Thus the contour   runs parallel to the imaginary
axis with a small positive real part, namely
$s=i\omega+\epsilon\,;\,-\infty \leq \omega \leq \infty$ with $\epsilon \rightarrow 0^+$.
 Therefore
\be G_k(t) = -\int \frac{d\omega}{2\pi } \frac{e^{i\omega\,t}
}{\Big[(\omega-i\epsilon)^2-\Ok^2- {\Sigma}_R(k,\omega) -i\Sigma_I(k,\omega)\Big]} \label{goft2} \ee which is recognized as the Fourier transform of the \emph{retarded} propagator,
where
\bea \Sigma_R(k,\omega) & = &  \frac{g^2}{2\pi}\, \int dk_0 \mathcal{P} \Big[\frac{\rho(k_0,k)}{\omega-k_0} \Big] \label{sigmaR}\\
\Sigma_I(k,\omega) & = & \frac{g^2}{2} \, \rho(\omega,k) \label{sigmaI}\,, \eea where $\mathcal{P}[\cdots]$ is the principal part and  $\Sigma_R(k,\omega), \Sigma_I(k,\omega)$ are even and odd functions of $\omega$ respectively as a consequence of the property (\ref{oddros}) which was used to write (\ref{sigmaR},\ref{sigmaI}).

In the non-interacting theory the integrand in (\ref{goft2}) features poles at $\omega^\pm = \pm \Ok$, including self-energy corrections the positions of the complex poles are determined by
\be \omega^2_p -\Ok^2-\Sigma_R(k,\omega_p)-i\Sigma_I(k,\omega_p)=0\,. \label{poles}\ee To leading order in $g^2$ the \emph{real} part of the poles are given by the renormalized frequencies
\be \omega^\pm_p = \pm \Okr ~~;~~ \Okr = \Ok+ \frac{\Sigma_R(k,\Ok)}{2\,\Ok}\,. \label{polerenor}\ee Near these poles and for small coupling, the propagator may be approximated   by writing $\omega = \omega_p + (\omega-\omega_p)$ and keeping the linear term in $(\omega-\omega_p)$ leading to the Breit-Wigner form of  the propagator in (\ref{goft2})
\be \frac{Z}{2\omega^\pm_p} \frac{1}{\Big[\omega-\omega^\pm_p-i\frac{\Gamma_k}{2}\Big]}\,, \label{BWprop}\ee where the wave function renormalizaton and decay width are given by
\bea Z^{-1} & = &  1- \frac{\Sigma'_R(k,\Ok)}{2\,\Ok} = 1+ \frac{g^2}{4\pi\Ok}\, \int dk_0 \mathcal{P} \Big[\frac{\rho(k_0,k)}{(\Ok-k_0)^2} \Big] \label{wavefuncren}\\
 {\Gamma_k}  & = & Z\,\frac{\Sigma_I(k,\Ok)}{{\Ok}} = \frac{g^2\,Z}{2\Ok}\,\rho(\Ok,k)  \,.\label{width}\eea  In (\ref{wavefuncren}) $\Sigma'_R = d\Sigma_R/d\omega$. Because $\Sigma_R,\Sigma_I$ are even and odd in $\omega$ respectively, it follows that $Z,\Gamma$ are the same for both poles.  The Fourier transform in (\ref{goft2}) is dominated by the complex poles of the Breit-Wigner propagator, we find
\be G_k(t)= \frac{Z}{\Okr}\,e^{-\frac{\Gamma_k}{2}t}\, \sin[\Okr t]\,. \label{solution}\ee

The Breit-Wigner approximation neglects the perturbatively small  contribution from the continuum along the branch cuts, describing the propagator in terms of simple resonances with complex poles near the real axis. The contribution from the continuum   gives rise to power law long time tails dominated by the behavior of the spectral density near the thresholds, with Landau damping being the   dominant contribution asymptotically for long time\cite{boylan}.

 Introducing this solution into (\ref{inhosolution}) it is   straightforward   to obtain the averages of  $\Psi_{\vk}(t)$ and  $\Psi_{\vk}(t)\Psi_{-\vk}(t)$ with the initial Wigner distribution function and the stochastic noise as in eqn. (\ref{totave}). We find
  \be \overline{\langle \langle \Psi_{\vk}(t)  \rangle \rangle } =  Z e^{-\frac{\Gamma_k}{2} t} \Big[ \overline{\Psi_{i,\vk}}~\cos[\Okr t] + \frac{\overline{\Pi_{i,\vk}}}{\Okr}~\sin[\Okr t] \Big] \,. \label{avePsi} \ee

 At $T=0$ the spectral density of the ``bath'' is given by (\ref{zeroTrho2chi}) and $Q^2 = m^2_{\phi,R} \ll (M_1+M_2)^2$  where $ m_{\phi,R} $ is the renormalized mass of the light $\phi$ field, in this case $\Gamma_k =0$ and only wave function and mass renormalization remain. After performing the average over the noise the remaining convolution integrals become similar to
 (\ref{t2int}) in the long time limit. Since $\Okr \ll (M_1+M_2)$ and only the two particle cut with threshold at $M_1+M_2$ remains in the spectral density at $T=0$, in this case ($T=0~;~\Gamma_k=0$)  we find that one power of $Z$ from the terms with $\Psi_i,\Pi_i$ are cancelled by the stochastic contribution and the long time limit yields
 \be \overline{\langle \langle \Psi_{\vk}(t)\Psi_{-\vk}(t) \rangle \rangle } =  \frac{Z}{2\Okr}\Big[1+\mathcal{O} \Big(\frac{g~\Okr}{M_1+M_2}\Big)^2\Big] \,. \label{2ptT0}\ee  In other words, the long time limit corresponds to the equal time two point correlation function of the field multiplied by wave function renormalization and with a renormalized frequency.  Thus we recover the case of the local effective field theory limit at $T=0$ emerging in the long-time limit after the transient dynamics has subsided, in complete agreement with the  discussion in section (\ref{sec:localqft}).

However, for $\Gamma_k\neq 0$ which is the case for $T\neq 0$, dissipative effects change dramatically the long time limit.

  For $\Gamma_k t \gg 1$ the contributions from the initial conditions vanish and only the stochastic part, namely the last term in (\ref{inhosolution})  yields a non-vanishing contribution, it is given by
   \be \Big(\frac{g\,Z}{2{\Okr}}\Big)^2~ \int \frac{dk_0}{(2\pi)}~ \frac{\rho(k_0,k)~\mathrm{coth}[\frac{k_0}{2T}]}{\Big[\big(\Okr-k_0\big)^2+\Big( \frac{\Gamma_k}{2}\Big)^2 \Big]} +\cdots \label{stoca}\ee where the dots stand for perturbatively small non-resonant contributions.

 In the narrow width approximation $\Gamma_k/\Okr \ll 1$ and using (\ref{width}) a straightforward calculation yields
 \be \overline{\langle \langle \Psi_{\vk}(t)\Psi_{-\vk}(t) \rangle \rangle } =  \frac{Z}{2\Okr}\,\big[1+ 2 n(\Okr) \big] \,. \label{thermal}\ee This is a remarkable result: up to the wave function renormalization factor $Z$ this is the equal time two point correlation function of a field in thermal equilibrium at temperature $T$ and for $T\gg \Ok$ it agrees with the classical result\cite{das}. This result is general regardless of the origin of the decay width $\Gamma$.

At $T\neq 0$ there are two new contributions to the spectral density (\ref{rho2chis}): the Landau damping term with a branch cut below the light cone $Q^2 \leq 0$, and the term $\rho_D$ with a branch cut for $0 < Q^2 < (M_1-M_2)^2$, their explicit expressions are given in appendix (\ref{sec:specdens}). The Landau damping term does not have support on the (renormalized) mass shell of the light field, as discussed in ref.\cite{boylan} this branch cut yields a long time power law tail and is perturbatively small in this case. The term $\rho_D$ yields a new phenomenon entirely. For $M_1> M_2 \gg m_{\phi_R}$ the spectral density has support on the mass shell of the light particle, therefore it determines the width $\Gamma_k$. This contribution to the spectral density describes the \emph{decay} process $\chi_1 \rightarrow \chi_2 \phi$ of the heaviest particle in \emph{the medium} and the inverse process. These processes lead to a \emph{damping} of the fluctuations of the $\phi$  and the \emph{thermalization} of the light particle $\phi$ with the heavy particles in the bath.  Thus the coupling of a light degree of freedom to a bath of heavy degrees of freedom that can \emph{decay} into the light field turns the $\phi$ particles into \emph{quasiparticles}\cite{boykev,drewes} with a finite lifetime that describes the thermalization of the latter. The decay of the heaviest environmental field builds the population of the light particle over a time scale $t_{th} \simeq 1/\Gamma$ until it thermalizes with the bath.

\subsection{Quantum kinetic interpretation of the $\phi$ width:}

The physical reason by which the \emph{decay} $\chi_1 \rightarrow \chi_2 \phi$ and the inverse process leads to thermalization of the light $\phi$ particle with the bath of heavy degrees of freedom  can be  understood from the  following quantum kinetic analysis.

The interaction Hamiltonian
\be H_I = g\int d^3 x \chi_1(x)\chi_2(x)\phi(x) \,,\label{Hikin}\ee with $M_1 > M_2 \gg m_\phi$  describes  the decay and inverse processes  in the medium
\be \chi_1 \leftrightarrow \chi_2 \phi \,.\label{decayinve}\ee The transition amplitude for the \emph{gain}  term of $\phi$ particles from the decay of the heavier particle in the medium, namely $\chi_1 \rightarrow \chi_2 \phi$ is given by
\be \mathcal{M}_{fi}\Big|_{\mathrm{gain}} = \frac{i g}{\sqrt{V}} ~ \frac{(2\pi)~\delta(w^{(1)}_k-w^{(2)}_{\vk+\vq}-\Omega_q)}{\sqrt{2w^{(1)}_k \,2w^{(2)}_{\vk+\vq}\,2\Omega_q}}~\sqrt{n_1(1+n_2)(1+N_\phi(q))}\,, \label{Mfigain} \ee  where $n_{1,2}$ are the equilibrium distribution functions for the heavy fields $\chi_{1,2}$
\be n_1 = n(w^{(1)}_k)~;~n_2 = n(w^{(2)}_p)~,~p=|\vq+\vk| \label{nchis}\ee
  and $N_\phi$ is the \emph{out of equilibrium} distribution for the light particles $\phi$.  The total gain probability per unit time is given by
\be \frac{dP_{\mathrm{gain}}}{dt} = \frac{g^2}{2\Omega_q} (2\pi) \int \frac{d^3k}{(2\pi)^3} ~\frac{\delta(w^{(1)}_k-w^{(2)}_{\vk+\vq}-\Omega_q)}{2w^{(1)}_k \,2w^{(2)}_{\vk+\vq}}~n_1(1+n_2)(1+N_\phi(q)) \label{dPdtgain}\ee The loss term from the inverse process $\chi_2 \, \phi \rightarrow \chi_1$ is obtained from the above expression by the replacement
\be n_1 \rightarrow (1+n_1)~;~1+n_2 \rightarrow n_2 ~;~ (1+N_\phi(q)) \rightarrow N_\phi(q)
\label{inverse}\ee leading to the quantum kinetic equation for the population of $\phi$ particles
\be \frac{dN_\phi(q)}{dt} = \frac{dP_{\mathrm{gain}}}{dt} -\frac{dP_{\mathrm{loss}}}{dt} \,, \label{gainminloss}\ee namely
\be \frac{dN_\phi(q)}{dt} = \frac{g^2\,\pi}{4\Omega_q}  \int \frac{d^3k}{(2\pi)^3} ~\frac{\delta(w^{(1)}_k-w^{(2)}_{\vk+\vq}-\Omega_q)}{w^{(1)}_k \,w^{(2)}_{\vk+\vq}}~
\Big[n_1(1+n_2)(1+N_\phi(q))-(1+n_1)n_2N_\phi(q) \Big]\,.\label{qkineq}\ee These processes are depicted in fig. (\ref{fig:kinetics}).

\begin{figure}[h!]
\includegraphics[height=3.5in,width=3.5in,keepaspectratio=true]{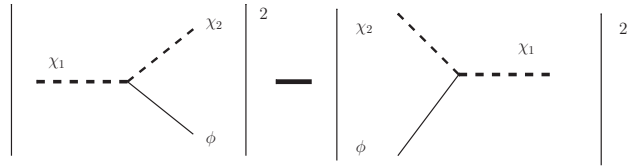}
\caption{Quantum kinetic interpretation of the decay width: the decay process $\chi_1\rightarrow \chi_2 \phi$ and its inverse lead to a build up of the population of $\phi$ particles in the medium.}
\label{fig:kinetics}
\end{figure}

This kinetic equation may be written as
\be \frac{dN_\phi(q,t)}{dt} = (1+N_\phi(q,t))\,\Gamma^<_q - N_\phi(q,t)\Gamma^>_q\,. \label{Gammaskin}\ee Inspection of (\ref{rho2gt}) along with the relation (\ref{relaqs}) shows that
$\Gamma^{>}_q$ is given by the second term in (\ref{rho2gt}) with $q_0 = \Omega(q)$ and $\Gamma^<_q$ is given by the
third term of $\rho^<(q0=\Omega_q,q)$. Furthermore it is straightforward to confirm that
\be \Gamma^>_q = e^{\beta \Omega_q} \Gamma^<_q\,\label{relagamas}\ee leading to the final form
 \be \frac{dN_\phi(q;t)}{dt} = -\gamma(q) \Big[N_\phi(q;t) - N_{eq}(q)\Big] \label{kinefin}\ee where
 \be N_{eq}(q) = \frac{1}{e^{\beta \Omega_(q)}-1}  \ee   and
 \be \gamma(q) = \Gamma^{>}_q -\Gamma^<_q = \frac{g^2}{2\Omega_q}~ \rho_D(q_0=\Omega_q,q)
\label{gamadist}\ee where $\rho_D(q_0=\Omega_q,q) $  is the second term of the spectral density of the bath (\ref{rho2chis}) evaluated  $q_0=\Omega_q$.
   The solution of (\ref{Gammaskin}) is
 \be N_\phi(q;t)) = N_{eq}(q)+ \Big[N_{\phi}(q;0)-N_{eq}(q)\Big]\,e^{-\gamma(q) t} \label{soluNfi}\ee

Since $\rho_D$ this is the only term in the spectral density with support at $q_0 = \Omega_q$  it follows that
\be \gamma(q) =  \frac{g^2}{2\Omega_q}~ \rho(q_0=\Omega_q,q) = \Gamma_q \Big(1+\mathcal{O}(g^2)\Big)\label{eqgamas}\ee where in the last equality we used the results (\ref{sigmaI},\ref{width}). Consequently,
\be \Gamma^<_q = \Gamma_q\,n(\Omega_q)~~;~~\Gamma^>_q = \Gamma_q\,\big[1+n(\Omega_q)\big]\,. \label{gams}\ee

This is one of the important results of this article: the decay of heavy fields in the medium into the light field leads to the thermalization of the light field with the heavy degrees of freedom. The dynamics of thermalization is non-unitary and is manifest in the dissipative kernels which are determined by the support of the spectral density of the bath on the mass shell of the light degree of freedom. This results in that the light field is described as a \emph{resonance} whose width is precisely the rate of approach to thermal equilibrium with the bath of heavy particles.

The equivalence between the results from the stochastic description and the quantum kinetic equation confirms the arguments of ref.\cite{carsten}   within a different framework and approach.

\subsection{$J[\phi]\neq \phi$: multiplicative noise}
The case $J[\phi]=\phi$ studied above highlights the stochastic nature of the effective action in a clear manner and exhibits the generalized fluctuation-dissipation relation between the self-energy of the light field and the correlation function of the stochastic noise. However the effective Langevin description is particular to the linear coupling to the bath.

For a generic coupling we now implement a Kramers-Moyal expansion in the relative coordinate\cite{calhubuk,schmid,hupaz}, writing $\phi^\pm$ in terms of the Wigner center of mass  ($\Psi$) and relative coordinates ($R$) as in (\ref{wigvars}),
\be J[\phi^\pm] = J[\Psi]\pm \frac{R}{2} J'[\Psi] + \frac{R^2}{8}\,J''[\Psi] +\cdots \label{KMexp}\ee Inserting this expansion into the influence action (\ref{Funravel}) and into the effective action (\ref{Leff}) we find
\begin{eqnarray}
iS_{eff}[\Psi,R] & = & \int_0^{\infty} dt \int d^3x \Big\{-i R(\vx,t)\left( \ddot{\Psi}(\vx,t)-\nabla^2 \,\Psi(\vx,t)-m^2_\phi \,\Psi(\vx,t)\right) \Big\} \nonumber \\
                 & - & \int_0^{\infty} dt \int d^3x \int_0^{\infty} dt' \int d^3x' \Big\{\frac{1}{2}\,R(\vx,t)J'[\Psi(\vx,t)]\,{\mathcal{N} }(\vx-\vx';t-t')\,R(\vx',t')J'[\Psi(\vx',t')]   \nonumber \\ & + &  R(\vx,t)J'[\Psi(\vx,t)]\,
i\Sigma^R(\vx-\vx';t-t')\, J[\Psi(\vx',t')]\Big\}  ~ + ~
 \int d^3x R_i(\vec x) \dot{\Psi}(\vec x,t=0)
\label{efflanwigJ}
\end{eqnarray}

Again, the term quadratic in   $R$ can be written in terms of a stochastic
noise as
\begin{eqnarray}
&& \exp\Big\{-\frac{1}{2} \int dt \int d^3x  \int dt' \int d^3x' R(\vx,t)J'[\Psi(\vx,t)]\,{\mathcal{N} }(\vx-\vx';t-t')\,R(\vx',t')J'[\Psi(\vx',t')] \Big\}  =  \nonumber \\  &&  \int {\cal D}\xi
\exp\Big\{-\frac{1}{2} \int dt \int d^3 x\int dt' \int d^3x' ~~ \xi(\vx,t)
\mathcal{N}^{-1}(\vx-\vx',t-t')\xi(\vx',t')
  + \nonumber \\ &&  i \int dt \int d^3 x~~ \xi(\vx,t) R(\vx,t)J'[\Psi(\vx,t)]\Big\}\,.
\label{noisefuncnew}
\end{eqnarray}
We   find that the exponential term in (\ref{genefunc}) is now given by
\bea && \exp\left\{-i \int_0^{\infty} dt \int d^3x~ ~ R(\vx,t) \left[
\ddot{\Psi}(\vx,t)-\nabla^2 \Psi(\vx,t)+m^2_\phi \Psi(\vx,t)+ \right. \right.\nonumber \\ && \left.\left.\int dt'\int d^3x' ~~ J'[\Psi(\vx,t)]\Sigma^{R}(\vx-\vx';t-t')J[\Psi(\vec x',t')]-J'[\Psi(\vx,t)]\,\xi(\vx,t) \right] \right\} \eea  where
\be{\cal P}[\xi]  =   \exp\left\{-\frac{1}{2} \int_0^{\infty} dt \int d^3 x
\int_0^{\infty} dt' \int d^3 x'~~ \xi(\vx,t) \mathcal{N}^{-1}(\vx-\vx',t-t')
\xi(\vx',t') \right\} \,. \label{probaxinew}\ee The Langevin equation of motion for $\Psi$ that follows from the effective action is now
\be
\ddot{\Psi}(\vx,t)-\nabla^2 \Psi(\vx,t)+m^2_\phi \Psi(\vx,t)+ J'[\Psi(\vx,t)]~ \int dt'\int d^3x' ~~ \Sigma^{R}(\vx-\vx';t-t')J[\Psi(\vx',t')]=J'[\Psi(\vx,t)]\,\xi(\vx,t)\,, \label{eqnofmotJpsi}\ee with
\be \langle \langle \xi(\vx,t) \rangle \rangle =0 ~~;~~ \langle \langle \xi(\vx,t)~\xi(\vx',t') \rangle \rangle =  \mathcal{N}(\vx-\vx',t-t')\,. \label{noisecor2}\ee
Therefore we see that when $J[\phi]\neq \phi$ we obtain also a stochastic description, but with  a multiplicative Gaussian noise.

From the time evolution of the Wigner function, or generating functional (\ref{genefunc}) we could implement the steps detailed in refs.\cite{leggett,schmid,grabert,hupaz,calhubuk} to obtain a Fokker-Planck equation, however the generating functional (\ref{genefunc}) in terms of the influence action when generalized to include external sources as in eqn. (\ref{generah}) contains already all the information necessary to obtain correlation functions at equal or different times by functional derivatives with respect to the sources. This is akin to the Martin-Siggia-Rose framework of stochastic classical phenomena\cite{msr}.

\section{The quantum Master equation:}\label{sec:master}
In the Master equation approach\cite{breuer,zoeller} the time evolution of the density matrix in considered in the interaction picture. With  the full density matrix $\hat{\rho}(t)$  given by eqn. (\ref{rhooft}) it follows that
\be \hat{\rho}_I(t)= e^{iH_0 t} \hat{\rho}(t) e^{-iH_0t}~~;~~H_0 = H_0[\phi]+H_0[\chi] \label{rhoIP}\ee whose time evolution obeys
\be \dot{\rho}_I(t) = -i \big[H_I(t),\hat{\rho}_I(t)\big] \label{rhodotip}\ee where $H_I(t)$ is the interaction Hamiltonian given by eqn. (\ref{hami}) in the interaction picture, namely the fields $\phi(\vec{x},t);\chi(\vec{x},t)$ evolve in time as free fields. The solution of (\ref{rhoIP}) is
\be \rho_I(t)= \hat{\rho}(0) -i \int^t_0 dt' \big[H_I(t'),\hat{\rho}_I(t')\big] \,. \label{solurhoip}\ee This solution is inserted back into (\ref{rhodotip}) leading to the iterative equation
\be \dot{\hat{\rho}}_I(t) = -i \big[H_I(t),\hat{\rho}(0)\big] -\int^t_0   \,\big[H_I(t),\big[H_I(t'),\hat{\rho}_I(t')\big]\big]\,dt' \,,\label{rhodotiter}\ee the next step relies on a series of \emph{assumptions}, the first being \textbf{: \underline{Factorization}}:
\be \hat{\rho}_I(t) = \hat{\rho}_{I\phi}(t)\otimes \rho_\chi(0)\,, \label{factorization}\ee namely that the heavy degrees of freedom (the ``bath'') remain in thermal equilibrium.  Taking the trace over the $\chi$ degrees of freedom an normal-ordering the interaction following eqn. (\ref{NO}) the first term on the right hand side of eqn. (\ref{rhodotiter}) vanishes upon taking the trace over the environmental degrees of freedom, and we find the evolution equation for the \emph{reduced density matrix} for $\phi$ in the interaction picture,
\bea \dot{\hat{\rho}}_{I\phi}(t)   & =  & -g^2\int^t_0 dt' \int d^3x \int d^3 x'\Bigg\{ \hat{J}(x) \, \hat{J}(x')\,\hat{\rho}_{I\phi}(t') \,\,G^>(x-x') + \hat{\rho}_{I\phi}(t') \,\hat{J}(x')\, \hat{J}(x)\,G^<(x-x') \nonumber \\
& - & \hat{J}(x) \,\hat{\rho}_{I\phi}(t') \, \hat{J}(x')\, G^<(x-x') - \hat{J}(x') \, \hat{\rho}_{I\phi}(t')\, \hat{J}(x)\,G^>(x-x') \Bigg\} \label{Linblad}\eea
where again $J(x) \equiv J[\phi(x)]$, and we use the shorthand convention $x \equiv (\vec{x},t)~;~x' \equiv (\vec{x}',t')$. The hat  on $J$ is to emphasize that these are (composite) operators in the interaction picture, $G^>, G^<$ are given by (\ref{ggreat},\ref{lesser}) and we suppressed the subscript $c$ (connected) because the operator $\mathcal{O}$ has been normal ordered. At this stage a \underline{\textbf{Markov}} approximation is usually invoked by replacing $\rho_{I\phi}(t') \rightarrow \rho_{I\phi}(t)$ taking it outside the time integral, this approximation is  justified in weak coupling. For example consider the first term in (\ref{Linblad}), it can be written as
\be -g^2 J(\vx,t) \int^t_0 \frac{d\mathcal{H}(t')}{dt'} \,\hat{\rho}_{I\phi}(t') \,dt' ~~;~~  \mathcal{H}(t') \equiv \int^{t'}_0 \hat{J}({\vx}',t'') \, G^>(\vx-{\vx}',t-t'')dt'' \label{incha}\ee which upon integration by parts yields
\be -g^2 J(\vx,t)  \mathcal{H}(t) \hat{\rho}_{I\phi}(t) + g^2  J(\vx,t) \int^t_0  \mathcal{H}(t') \,\frac{d\hat{\rho}_{I\phi}(t')}{dt'} dt' \label{incha2}\ee in the second term $d\hat{\rho}_{I\phi}(t')/dt' \propto g^2$ so this   term yields a contribution that is formally of order $g^4$ and can be neglected to second order. The same analysis can be applied to all the other terms in (\ref{Linblad}) with the conclusion that in weak coupling and  to leading order $(g^2)$  the Markovian approximation  $\hat{\rho}_{I\phi}(t') \rightarrow \hat{\rho}_{I\phi}(t)$ is justified.

Therefore in the Markov approximation the quantum master equation becomes
\bea \dot{\hat{\rho}}_{I\phi}(t)   & =  & -g^2\int^t_0 dt' \int d^3x \int d^3 x'\Bigg\{ \hat{J}(x) \, \hat{J}(x')\,\hat{\rho}_{I\phi}(t) \,\,G^>(x-x') + \hat{\rho}_{I\phi}(t) \,\hat{J}(x')\, \hat{J}(x)\,G^<(x-x') \nonumber \\
& - & \hat{J}(x) \,\hat{\rho}_{I\phi}(t) \, \hat{J}(x')\, G^<(x-x') - \hat{J}(x') \, \hat{\rho}_{I\phi}(t)\, \hat{J}(x)\,G^>(x-x') \Bigg\} \,. \label{markovlim}\eea

The relation between this master equation in operator form and the effective action of the previous section which is cast in terms of functional integrals in the field basis  is not \emph{a priori} evident. However we can infer the equivalence from the result given by eqn. (\ref{dtF}). The first step is to recall that the variables with $+$ correspond to the forward time evolution branch, namely from the time-ordered evolution, whereas those with   $-$ correspond to the backward time evolution branch, namely, anti-time ordered evolution. Because the time evolution of the density matrix is determined by $U(t)\rho(0)U^{-1}(t)$ with $U(t)$ the forward time evolution operator (time ordered) and $U^{-1}(t)$ the backward time evolution operator (anti-time ordered),  operator insertions with $+$ go \emph{before the density matrix}, and operator insertions with $-$ go \emph{after} the density matrix. Therefore when multiplying the density matrix the equivalence becomes
\be A^+ B^+ \rightarrow A B \rho ~~;~~A^- B^- \rightarrow \rho A B~~;~~ A^+ B^- \rightarrow  A \rho B~~;~~ A^- B^+ \rightarrow B \rho A \label{equivalence}\ee Therefore this equivalence applied to  (\ref{markovlim})   states that the first term in (\ref{markovlim}) is identified with $J^+(x) J^+(x') G^>(x-x') $, the second with  $ J^-(x) J^-(x') G^<(x-x') $, the third with $- J^+(x) J^-(x')  G^<(x-x')$ and the fourth with $-J^-(x) J^+(x')G^>(x-x')$. From these identifications  we immediately recognize that these are \emph{precisely } the terms in the \emph{time derivative} of the influence function (\ref{dtF}), which describes the \emph{interaction} of the light particle with the bath of heavy fields. This ``dictionary'' establishes the direct correspondence between the influence function and the master equation approaches: \emph{the solution of the quantum master equation (\ref{Linblad}) in the field basis and in the Markov approximation is the influence action (\ref{Funravel})}.

 The factorization assumption (\ref{factorization}) is equivalent to obtaining the influence action from the correlation functions of the bath in the initial density matrix (of the bath) and exponentiating the result in a cumulant expansion, as manifestly exhibited in the result (\ref{trasa}, \ref{expec})   for the influence action.

\subsection{Local effective field theory: effective Hamiltonian.}
In order to understand how a local effective field theory emerges from the quantum master equation, and to establish contact with the results of section (\ref{sec:localqft}), let us consider the interaction $g J[\phi]\,\chi$ with only one heavy field ($\chi$)  of mass $M \gg m_\phi$ at $T=0$. Writing the correlation functions $G^\lessgtr$ as in eqn. (\ref{Ggfd},\ref{Glfd}) with the  spectral densities   given by (\ref{rhosis}), defining
\be \int d^3 x J[\vx,t] e^{i\vk\cdot \vx} = \int dp_0 e^{-ip_0t}\widetilde{J}(\vk,p_0) \,, \label{transf}\ee and carrying out the integral in $t'$ in (\ref{markovlim}) in the long time limit $t\rightarrow \infty$ with a convergence factor $\epsilon \rightarrow 0^+$, we find
\bea \dot{\hat{\rho}}_{I\phi}(t)   & =  &  -i \,g^2 \int \frac{d^3k}{(2\pi)^3\,2 w_k}\int dp_0 \int dq_0\,e^{-i(q_0+p_0)t}~\Bigg\{ \frac{\widetilde{J}(\vk,p_0) \widetilde{J}(-\vk,q_0)~  \hat{\rho}_{I\phi}(t) }{(q_0-w_k +i\epsilon)} +  \frac{\hat{\rho}_{I\phi}(t)\,\widetilde{J}(\vk,q_0) \widetilde{J}(-\vk,p_0) }{(q_0+w_k +i\epsilon)}  \nonumber \\ &- &   \frac{\widetilde{J}(\vk,p_0)~\hat{\rho}_{I\phi}(t)  ~ \widetilde{J}(-\vk,q_0)}{(q_0+w_k +i\epsilon)}- \frac{\widetilde{J}(\vk,q_0)~\hat{\rho}_{I\phi}(t)  ~ \widetilde{J}(-\vk,p_0)}{(q_0-w_k +i\epsilon)} \Bigg\}\,. \label{localqme} \eea If the frequency $q_0$ and momentum $k$ transfer are $q_0, k \ll M$ we can set $q_0 \simeq 0$ in the denominators and replace $w_k = M$. In this low energy limit the last two terms in (\ref{localqme}) cancel each other out, and from (\ref{transf}) we obtain
\be  \dot{\hat{\rho}}_{I\phi}(t)    =    -i\,\big[H^{I}_{eff}(t),\hat{\rho}_{I\phi}(t) \big] \label{hefqme}\,, \ee where
\be   H^{I}_{eff}(t)     =      -\,\frac{g^2}{2M^2}\int d^3x J^2[\phi(\vx,t)]\,. \label{effhami} \ee
Therefore we recover the local ``current-current'' limit with Hamiltonian evolution for the reduced density matrix in the interaction picture in complete agreement with the result (\ref{heffunitary}) from the influence action approach. This result confirms that obtained from the influence action in section (\ref{sec:localqft}). In particular, the cancellation of the last two terms in (\ref{localqme}) in the ``local limit'' is the same as the cancellation of the last two terms in (\ref{Fham}),  furthermore in this local limit the limit $\epsilon \rightarrow 0$ can be taken safely without yielding a delta function, such term would yield a purely real, namely dissipative, contribution to the quantum master equation. The cancellation of the last two terms and the vanishing of the dissipative contribution in agreement with the results of section (\ref{sec:localqft}) are a direct manifestation of the equivalence between the influence action and the quantum master equation discussed above.

 \subsection{$J[\phi]=\phi$ Lindblad Master equation.}

 The equivalence with the influence function and the results of the stochastic Langevin equation in particular the dynamics of damping and thermalization become  more clear by studying the   case   $J[\phi] = \phi$ to compare with the results of section (\ref{sec:langevin}). In the interaction picture and in terms of the spatial Fourier transform in a volume $V$

\be \phi(\vx,t)= \frac{1}{\sqrt{V}} \sum_{\vq} \phi_{\vq}(t) e^{i\vq\cdot\vx} \label{fiq}\ee  where \be \phi_{\vq}(t) = \frac{1}{\sqrt{2\Omega_q}}\Big[a_{\vq}\,e^{-i\Omega_q t} + a^\dagger_{-\vq} \, e^{i\Omega_q t} \Big] \label{aadag}\ee and the operators $a_{\vq};a^\dagger_{-\vq}$ do not depend on time. At this point we invoke yet another approximation: the \textbf{``rotating wave approximation''}:  in writing the products $\phi(\vx,t)~\phi(\vx',t')$ in (\ref{Linblad}) there are two types of terms  with very different time evolution. Terms of the form
\be a^\dagger_{\vq}~a_{\vq}~e^{i{\Omega_q(t-t')}} \label{ada} \ee and its hermitian conjugate are ``slow'', and terms of the form
\be a^\dagger_{\vq}~a^\dagger_{-\vq}~ e^{2i\Omega_q t}~e^{i{\Omega_q(t-t')}}~~;~~ a_{\vq}\,a_{-\vq} e^{-2i\Omega_q t}\,e^{-i{\Omega_q(t-t')}} \label{adad}\ee are fast, the extra rapidly varying phases lead to rapid dephasing and do not yield resonant (energy conserving) contributions. These terms only give perturbatively small transient contributions and are discussed below. Keeping only the slow terms which dominate the long time dynamics and neglecting the fast oscillatory terms defines the ``rotating wave approximation'' ubiquitous in quantum optics. We will adopt this approximation and comment later on the corrections associated with keeping the fast   terms. Implementing the \textbf{Markov}  approximation $\rho_{I\phi}(t')\rightarrow \rho_{I\phi}(t)$, and the   \textbf{rotating wave } approximation (keeping only terms of the form $a^\dagger~a, a~a^\dagger$) using the spectral representation of the correlators (\ref{specrep}) with the property $\rho^<(-k_0,k)=\rho^>(k_0,k)$  and carrying out the spatial and temporal integrals we obtain the  \emph{Lindblad} form of the quantum master equation,
\bea \dot{\hat{\rho}}_{I\phi}(t)  & = &  \sum_{\vk} \Bigg\{   -iR_k(t)~\Big[a^\dagger_{\vk}\,a_{\vk}, \hat{\rho}_{I\phi}(t) \Big] \nonumber \\
& - & \frac{\Gamma^>_k(t)}{2} \Big[a^\dagger_{\vk}\,a_{\vk}~ \hat{\rho}_{I\phi}(t) + \hat{\rho}_{I\phi}(t)~a^\dagger_{\vk}\,a_{\vk} - 2 a_{\vk}~\hat{\rho}_{I\phi}(t)~a^\dagger_{\vk} \Big]\nonumber \\
& - & \frac{\Gamma^<_k(t)}{2} \Big[a_{\vk}\,a^\dagger_{\vk}~ \hat{\rho}_{I\phi}(t) + \hat{\rho}_{I\phi}(t)~a_{\vk}\,a^\dagger_{\vk} - 2 a^\dagger_{\vk}~\hat{\rho}_{I\phi}(t)~a_{\vk} \Big] \Bigg\} \,,
\label{Linfin} \eea  where
\be R_k(t) = \frac{g^2}{4\pi\Ok} \,\int dk_0 \rho(k_0,k)\,\frac{\Big[1-\cos[(\Ok-k_0)t] \Big]}{(\Ok-k_0)} \label{Roftim}\ee
\be \Gamma^>_k(t) = \frac{g^2}{2\Ok} \,\int dk_0 \rho(k_0,k)\,\big[1+n(k_0)\big]\frac{ \sin[(\Ok-k_0)t] }{\pi(\Ok-k_0)} \label{gamgre}\ee
\be \Gamma^<_k(t) = \frac{g^2}{2\Ok} \,\int dk_0 \rho(k_0,k)\, n(k_0) \frac{ \sin[(\Ok-k_0)t] }{\pi(\Ok-k_0)} \label{gamles}\ee
The second and third lines in (\ref{Linfin}) are called the \emph{dissipator}, these are non-Hamiltonian, purely dissipative terms that cannot be written in terms of a local hermitian effective field theory.

Taking the long time limit at this stage using the results on the right hand side of eqn. (\ref{t2int}), would lead to
\be R(t) ~~\overrightarrow{t\rightarrow \infty} ~~ \frac{g^2}{4\pi\Ok} \,\int dk_0 \,\mathcal{P}\Bigg[\frac{\rho(k_0,k)}{ \Ok-k_0}\Bigg] = \frac{\Sigma_R(\Ok,k)}{2\,\Ok} =  \delta \Ok \,, \label{renfreqlin} \ee which is recognized as the renormalization of the frequency from eqns. (\ref{sigmaR},\ref{polerenor}). Similarly in the long time limit
\be \Gamma^>_k(t) ~~\overrightarrow{t\rightarrow \infty} ~~   \frac{g^2}{2\Ok} \, \rho(\Ok,k)\,\big[1+n(\Ok)\big] = \Gamma^>_k \label{gamaglontim}\ee
\be \Gamma^<_k(t) ~~\overrightarrow{t\rightarrow \infty} ~~   \frac{g^2}{2\Ok} \, \rho(\Ok,k)\, n(\Ok)  = \Gamma^<_k \label{gamallontim}\ee
\be \Gamma^>_k(t)-\Gamma^<_k(t) ~~\overrightarrow{t\rightarrow \infty} ~~   \frac{g^2}{2\Ok} \, \rho(\Ok,k)   = \Gamma_k \label{gamadiflontim}\ee
where $\Gamma^{\lessgtr}_k,\Gamma_k$ are \emph{precisely} the rates in the quantum kinetic equation (\ref{Gammaskin},\ref{gams}) and coincide to leading order in $g^2$ with the result in eqn. (\ref{width}). Therefore in the long time limit the first term in the quantum master equation (\ref{Linfin}) is identified as a \emph{local} contribution to the effective Hamiltonian (in quantum optics this term is referred to as the Lamb shift), and the second and third contributions correspond to the dissipative terms in the Lindblad master equation. In particular at $T=0$ the long time limit leading to (\ref{gamaglontim},\ref{gamallontim}) would vanish because at $T=0$ the spectral density does not have support on the mass shell of the light field, and features a two particle threshold at $|k_0| = \sqrt{k^2+(M_1+M_2)^2} >> \Ok$. Therefore we will proceed to analyze the consequences of the Lindblad master equation (\ref{Linfin}) without taking the long time limit at this stage.

For any interaction picture operator $\mathcal{A}$ associated with the light field $\phi$
\be \frac{d}{dt}\langle \mathcal{A}\rangle = \mathrm{Tr}\Big\{ \dot{\mathcal{A}}~\hat{\rho}_{I\phi}(t) + {\mathcal{A}}~\dot{\hat{\rho}}_{I\phi}(t)\Big\} \,, \label{timederave}\ee where the average $\langle (\cdots) \rangle = \mathrm{Tr}(\cdots)\hat{\rho}_{I\phi}(t)$. Since $a_{\vk} ~,~ a^\dagger_{\vk}$ are time independent in the interaction picture, we find
\bea \frac{d}{dt}\langle a_{\vk} \rangle  & = &  \Big[-iR(t)-\frac{\Gamma_k(t)}{2} \Big]\, \langle a_{\vk} \rangle \nonumber \\\frac{d}{dt}\langle a^\dagger_{\vk}\rangle  & = &  \Big[iR(t)-\frac{\Gamma_k(t)}{2} \Big]\, \langle a^\dagger_{\vk} \rangle \label{aveaad}\eea where
\be \Gamma_k(t) = \Gamma^>_k(t)-\Gamma^<_k(t)\,. \label{totigam}\ee Using the asymptotic integrals
 \be t\int_{-\infty}^{\infty} dk_0 \, \frac{\rho(k_0,k)}{( \Ok-k_0)}\,\Bigg[ 1-\frac{\sin(\Ok-k_0)\,t}{(\Ok-k_0)\,t} \Bigg] ~~\overrightarrow{t\rightarrow \infty} ~~ t\, \int_{-\infty}^{\infty} dk_0 \, \mathcal{P}\Big[\frac{\rho(k_0,k)}{(\Ok-k_0)}\Big] \label{realpartofE}\ee
 \be \int_{-\infty}^{\infty} dk_0 \, \frac{\rho(k_0,k)}{( k_0-\Ok)^2}\,\Bigg[ 1-\cos\big[(k_0-\Ok)t\big] \Bigg]~~ \overrightarrow{t\rightarrow \infty} ~~\pi\,t\, \rho(\Ok,k) + \int_{-\infty}^{\infty} d\omega' \,\mathcal{P}\Big[ \frac{\rho(k_0,k)}{( \Ok-k_0)^2} \Big] \label{imagE}\ee we find in the long time limit
 \be \int^{t\rightarrow \infty}_0 R(t')dt'  = \delta \Omega_k~t \label{Rlontim}\ee
\be \int^{t\rightarrow \infty}_0 \frac{\Gamma_k(t')}{2}\,dt'  = \frac{\Gamma_k}{2}~t + Z^{-1}-1 \label{gamlontim}\ee where $Z^{-1}$ is given by eqn. (\ref{wavefuncren}). The emergence of the wave function renormalization is a consequence of keeping the time dependence in $\Gamma^\lessgtr (t)$ in (\ref{Linfin}) until the full solution is obtained. Taking the long-time   limit too hastily in the Lindblad equation does not allow to extract the wave function renormalization asymptotically.

Therefore in the long time limit we find
\be \langle a_{\vk}\rangle(t) \simeq  Z\, \langle a_{\vk}\rangle(0)\,e^{-i\delta \Omega_k~t}~e^{-\frac{\Gamma_k}{2}~t} ~~;~~ \langle a^\dagger_{\vk}\rangle(t) \simeq  Z\, \langle a^\dagger_{\vk}\rangle(0)\,e^{i\delta \Omega_k~t}~e^{-\frac{\Gamma_k}{2}~t}\,. \label{asyaada}\ee Introducing these solutions into $\langle \phi_{\vk}(t) \rangle $ by replacing  $\langle a_{\vk} \rangle;\langle a^\dagger_{\vk} \rangle$ into  (\ref{fiq})  and recognizing that $\Ok+\delta \Ok =\Okr$ we find in the long time limit
\be \langle \phi_{\vk}(t) \rangle = Z\,e^{-\frac{\Gamma_k}{2} t}~ \Bigg[\langle \phi_{\vk}(0) \rangle \cos[\Okr t] ~+~ \frac{\langle \pi_{\vk}(0)\rangle}{\Okr} \sin[\Okr t]\Bigg] ~~;~~ \langle \pi_{\vk}(0)\rangle = \frac{d}{dt} \langle \phi_{\vk}(t) \rangle\Big|_{t=0} \,,\label{avefiqm}\ee this long time solution is the same as that obtained with the influence functional given by eqn. (\ref{avePsi}).

To establish the relation with the influence action result (\ref{thermal}) we now consider the evolution equation for averages of bilinears:
\be \frac{dN_k(t)}{dt} = -\Gamma_k(t) N_k(t) + \Gamma^<_k(t)~~;~~ N_k(t) = \langle a^\dagger_{\vk}a_{\vk} \rangle \,, \label{dNdt}\ee
\bea \frac{d}{dt} \langle a_{\vk}~ a_{-\vk} \rangle & = &  \Big[-2iR_k(t) - \Gamma_k(t)\Big] \langle a_{\vk}~ a_{-\vk} \rangle \nonumber \\
 \frac{d}{dt} \langle a^\dagger_{\vk}~ a^\dagger_{-\vk} \rangle & = &  \Big[2iR_k(t) - \Gamma_k(t)\Big] \langle a^\dagger_{\vk} ~ a^\dagger_{-\vk} \rangle\, , \label{bilins}\eea
 Taking the long time limit in $\Gamma^\lessgtr_k(t) \rightarrow \Gamma^\lessgtr_k$ the rate equation (\ref{dNdt}) agrees with eqn. (\ref{kinefin}) from quantum kinetics, neglecting the (perturbative) contribution from wave function renormalization, the solution of (\ref{dNdt}) in the long time limit  is given by
\be N_k(t) = N_{eq,k}+\big[N_k(0)-N_{eq,k}\big]\,e^{-\Gamma_k t}\label{thermalasy}\ee which describes thermalization when $M_1>M_2 \gg m_\phi$, namely when (at least) one of   the heavy fields in the medium can decay into the light field, in agreement with the influence action, stochastic and  quantum kinetics approaches   to leading order in $g^2$. Furthermore, the  solutions to the equations (\ref{bilins})  for the bilinears become in the long time limit
\be \langle a_{\vk} ~ a_{-\vk} \rangle (t) = \langle a_{\vk} ~ a_{-\vk} \rangle (0)~ e^{-2iR_k t}~e^{-\Gamma_k t} \,, \label{solubilins} \ee and its hermitian conjugate, again neglecting wave function renormalization perturbative contributions. Combining these results we find in the long time limit
\be \langle \phi_{\vk}(t) \phi_{-\vk}(t) \rangle \rightarrow \frac{1}{2\Ok}\Big[1 + 2 N_{eq,k}\Big] \,, \label{thermalfi2}\ee which again exhibits a stationary thermal correlation function in agreement with the results of the influence action to leading order in $g^2$. The renormalization of the frequency $\Ok \rightarrow \Okr$ can be included by writing the free field theory of the light field $\phi$ in terms of the renormalized mass and introduce a counterterm in the perturbation, in the long time limit the Hamiltonian term $R_k(t)$ in the Lindblad equation will cancel the frequency counterterm.

The solutions (\ref{thermalasy}) and (\ref{thermalfi2}) emerges in the long time limit provided the spectral density has support on the mass shell of the light field $\phi$ namely if the decay process $\chi_1 \rightarrow \chi_2 \phi$ is available.  At $T=0$ for example the spectral density has a two particle threshold above the $\phi$ mass shell and $\Gamma_k =0$. In this case there are no resonant terms, if the long time limit is taken (hastily) in $\Gamma^\lessgtr_k(t)$ (\ref{gamgre},\ref{gamles}) then the dissipative contribution to the Lindblad master equation would vanish identically.

However, keeping the time dependence of $\Gamma^\lessgtr_k(t)$ reveals important aspects that are missed if the long time limit is taken  at the outset, these emerge from the solution of the rate equation (\ref{dNdt}) with the time dependent rates. It is given by
\be N(t)=e^{-\int^t_0 \Gamma_k(t')dt'}~\Bigg[N(0)+\int_0^t \Gamma^<_k(t')~e^{\int^{t'}_0 \Gamma_k(t'')dt''} \,dt' \Bigg] \,. \label{noftnosec}\ee The long time asymptotic behavior is obtained from the identities (\ref{gamlontim}), if $\Gamma_k\neq 0$ the secular term dominates at long times and memory of the initial conditions is lost after a time $1/\Gamma_k$ leading to the result in eqn. (\ref{thermalasy}).

On the other hand, in the case when $\Gamma_k=0$ as in $T=0$, the long time limit of the time integrals are non-secular and given by (\ref{wavefuncren}), where $Z^{-1}-1 \propto g^2$, therefore to leading order in $g^2$ we can neglect the exponential terms in (\ref{noftnosec}) and assuming that the field $\phi$ is initially in the ground state, namely $N(0)=0$ we find that asymptotically
\be N(t\rightarrow \infty) \propto Z^{-1}-1  \propto g^2 \,.\label{asynoftnosec}\ee This result has a clear interpretation: after the transient dynamics subsides the coupling to the bath of heavy particles has produced virtual excitations of the $\phi$ field, and the ``true'' (or dressed) ground state is a linear superposition of the ``bare'' ground state and multiparticle excitations. The wave function renormalization measures the overlap between the bare and the dressed states and the number of ``bare'' particles in the ``dressed'' state.

An important conclusion from the equivalence with the stochastic description is the identification between the ``rotating wave approximation'' and the Breit-Wigner approximation for the propagator (\ref{BWprop}) which replaces the full propagator by a complex pole which describes reliably the long time limit.

\vspace{2mm}

\textbf{Counterrotating terms:}

\vspace{1mm}

In the derivation of the Lindblad master equation (\ref{Linfin}) we  neglected terms of the form
\be a_{\vk}\,a_{-\vk}\,e^{-2i\Ok t} e^{i\Ok(t-t')} ~~;~~ a^\dagger_{\vk}\,a^\dagger_{-\vk}\,e^{2i\Ok t} e^{-i\Ok(t-t')}\,. \label{counter}\ee The time integral over $t'$ can be carried out as in the Lindblad form yielding contributions of the form
$a_{\vk}\,a_{-\vk}\,e^{-2i\Ok t} \rho^\lessgtr(k_0,k) \hat{\rho}_{I\phi}(t)$ etc. The contribution of these terms to the equations of motion for linear or bilinear forms of $a,a^\dagger$ are straightforward to obtain, they do not yield terms that grow secularly in time because the rapid dephasing of the oscillatory terms average out in the time integrals. A simple analysis shows that these terms yield perturbatively small subleading contributions of the form $\delta \Ok/\Ok~;~\Gamma_k/\Ok$ as compared to those obtained from the Lindblad form which captures the secular growth in time because of the resonances and describes the leading behavior in the long time dynamics.

In conclusion, the ``counterrotating'' terms   always yield perturbatively small contributions that are bound in time. They are negligible in the case when the spectral density of the bath has support on the mass shell of the light field (resonant terms), and are perturbatively small in the case when it does not, in agreement with the perturbative corrections obtained from the influence action approach in this case, (see eqn. (\ref{2ptT0})).

\section{Discussion}
We have established the relation between the quantum master equation and the influence action and compared the evolution of expectation values and correlation functions from the two approaches. Each approach has advantages and disadvantages that merit a discussion.

\begin{itemize}
\item{\textbf{Influence action}: the effective action (\ref{Leffhs}) and the generating functional (\ref{generah}) yield the time evolution of the density matrix for the system in the presence of external sources and functional derivatives with respect to these sources yield any arbitrary correlation function. The stochastic nature of the time evolution is more explicitly manifest in this approach and the correlations of the bath enter in a more explicit manner in the statistical and stochastic averages. An important advantage of the influence action is that it allows to obtain correlation functions at \emph{different times} from functional derivatives with respect to the external sources.  }

 \item{\textbf{Quantum master equation and Lindblad form:} This approach is more useful to obtain directly the time evolution of \emph{slow}  averages, namely averages of operators that do not evolve in time in the interaction picture. The connection with the quantum kinetic description is more readily recognized in this approach as is manifest in the results (\ref{dNdt}, \ref{noftnosec}). The general time evolution of averages of operators is given by eqn. (\ref{timederave}), the first term of which requires the time evolution of the density matrix, not just its time derivative. Therefore in order to extract the time evolution of averages of arbitrary operators it is necessary to obtain the full time evolution of the density matrix, only the time evolution of averages associated with operators that do not depend on time in the interaction picture are simple(r) to obtain. Correlation functions at different times are not straightforward to obtain from the quantum master equation either in Lindblad form or the more general form including the counterrotating terms. The relationship between the general form of the quantum master equation (\ref{Linblad}) discussed in section (\ref{sec:master}) and the \emph{time derivative} of the influence function (\ref{dtF}) clarifies that the influence action \emph{is the solution} of the quantum master equation for the reduced density matrix in the interaction picture in the field basis. Furthermore this equivalence also sheds light on the nature of the factorization assumption in the master equation formulation.  Therefore the effective action (\ref{Leff}) and in particular the generating functional in terms of the Wigner transform (\ref{deltaprob}) give the time evolution of the reduced density matrix.}

 \item{\textbf{Thermalization:} In our study thermalization emerges as a consequence of the decay of a heavy particle in the bath into another heavy particle of smaller mass and a light particle. This case is actually quite general:  in principle one would expect that the application of effective actions to low energy physics (or long time, long wavelength phenomena) is a result of tracing over a large number of heavy degrees of freedom of various masses and couplings to the light degrees of freedom. Therefore   the particular case in this study is, quite likely, representative of the general situations in which an effective low energy description is sought. When the temperature of the environment is comparable to the energy scale of the heavy degrees of freedom, these will be excited and present with large population in the plasma. However, thermalization of the light degree of freedom will occur even when the temperature of the environment is  \emph{much smaller} than  the masses of the heavy fields. Obviously the population of these fields will be exponentially suppressed, and this is reflected in the decay rate $\Gamma_k$ through the spectral representation (see appendix \ref{sec:specdens}), however,  this entails   that the time scale to full thermalization $\tau \simeq 1/\Gamma$ becomes very long, nevertheless thermalization will ensue and the light ``system'' will eventually reach a stationary, thermal state on this time scale. The non-perturbative nature of this result is noteworthy: if the temperature is much smaller than the masses of the heavy fields, the time scale to thermalization is very long and the secular growth of the fluctuations become very large. In the long time limit even for small coupling and large masses, the \emph{fluctuating} part of two point correlation function of the light field at equal time approaches the thermal distribution (\ref{thermalfi2}).      As discussed in detail in the above sections, this process of thermalization is necessarily described by the non-local  dissipative contributions to the effective action.  }

\end{itemize}

\section{Conclusions and further questions. }\label{sec:conc}
In this article we studied the emergence of an effective field theory out of equilibrium from the open quantum system perspective where a light field $\phi$ of mass $m_\phi$ -- the ``system''-- is  in interaction with a bath or environment at temperature $T$, taken to be either one scalar heavy field $\chi$ with mass $M\gg m_\phi$ or two scalar  heavy fields $\chi_1,\chi_2$ with a hierarchy of masses $M_1>M_2\gg m_\phi$. We obtain the reduced density matrix of the field $\phi$ by tracing out the environmental degrees of freedom up to second order  for interactions of the form $J[\phi]\,\mathcal{O}[\chi_j]$ where $J,\mathcal{O}$ are in general polynomials of the fields. The time evolution of the reduced density matrix is determined by the \emph{influence action} which is generally non-local and describes dissipative processes and non-unitary time evolution of the reduced density matrix.
For the case when the interaction is of the form $g \,J[\phi] \,\chi$ we show how the familiar ``current-current'' effective field theory emerges from the influence function in the low frequency, long wavelength limit at long times after the transient dynamics subsides.

When the environment contains \emph{two} heavy fields $\chi_1,\chi_2$  the spectrum of environmental fluctuations   is very rich and leads to a wealth of dynamical phenomena.  For an interaction $g \phi \chi_1\chi_2$ a Wigner (semiclassical) transform of the effective action yields a stochastic description of the dynamics described by a Langevin equation with   non-local dissipative and additive  Gaussian noise kernels that obey a generalized fluctuation dissipation relation. When the interaction is non-linear in  $\phi$    the noise is multiplicative.
 At $T=0$ the spectral density of the bath   only features a two particle threshold well above the mass shell of the light field and the influence action of the bath yield  mass and  wave function renormalizations, the asymptotic long time behavior corresponds to a \emph{renormalized} local effective field theory.

For $T \neq 0$ new ``anomalous'' thresholds emerge and the spectral density of the bath features support on the mass shell of the light field that describes the \emph{decay} $\chi_1 \rightarrow \chi_2 \phi$. This ``in medium'' process is manifest as non-local dissipative contributions to the effective action which ultimately lead to the \emph{thermalization} of the light field with the bath.  Even for $T\ll M_1,M_2$ when the population of heavy degrees of freedom is thermally suppressed, the light field thermalizes at long times. We show that the dissipative contributions and the thermalization of the light field from the decay of the heavy fields in the bath  are in complete agreement with    quantum kinetic results.  We argue that this result is quite general in that an environment with heavy degrees of freedom will include a hierarchy of very massive states which upon interacting with the light fields will result in their decay into the light degrees of freedom. This decay necessarily implies that the spectral density of the bath will have support on the mass shell of the light particle, which in turn leads to a dissipative contribution to the effective action and eventual thermalization of the light field.

We also obtain the quantum master equation up to second order in the interaction and show directly that its solution in the field basis is precisely the\emph{ influence action} and elucidate the nature of the various approximations invoked in the quantum master equation approach. For the case $g\phi\chi_1\chi_2$ we obtain the Lindblad form of the quantum master equation  under precise approximations which are also understood from the equivalence with the influence action. It contains a Hamiltonian and a dissipative term whose coefficients depend explicitly on time. We show that this time dependence is crucial to obtain consistently important renormalization aspects.  The asymptotic time evolution obtained from the Lindblad master equation with the time dependent coefficients is in complete agreement with that obtained from the influence action,  the stochastic Langevin equation  and the results from  quantum kinetics.

While the influence function, Langevin and quantum master equation approaches all agree in the long time limit, they offer advantages and disadvantages which we discussed in detail. In particular, the influence action approach leads directly to the stochastic Langevin description, while the stochastic nature is not \emph{a priori} evident in the quantum master equation. Furthermore, the infuence action approach when augmented to include external sources allows to obtain correlation functions at \emph{different times} which is more difficult in the master equation approach.

These results offer a note of caution on the application of effective field theories in a finite temperature (and likely a finite density) environment  such as the early Universe, as dissipative effects arising from the influence of heavy environmental fields lead to non-local and non-unitary stochastic dynamics of the light degrees of freedom.

\vspace{2mm}

\textbf{Further questions:}

The time evolution of the initially prepared density matrix leads to the \emph{entanglement} of the system and environment degrees of freedom. After tracing over the environment degrees of freedom, the influence action describes the time evolution of the reduced density matrix and is a direct result of the entanglement between the system and the environment. The Von-Neumann entropy associated with the reduced density matrix is the entanglement entropy (although alternative definitions of the entanglement entropy are available\cite{daley}). Several  questions   remain  to be explored: a) how to extract the entanglement entropy from the influence action, b)  whether the time evolution of the entanglement entropy has a quantum kinetic interpretation, as for example an $H$-theorem, c) what are the manifestations of the stochastic nature of the effective action upon the entanglement entropy.

Our ultimate goal is to study the emergence of effective field theories in cosmology, in particular during the inflationary stage,  under the assumption that there are heavy degrees of freedom with mass or energy scales larger than the Hubble scale that are traced over leading to an effective description in terms of a single ``light'' scalar field. In inflationary cosmology there are novel processes associated with the lack of a global time-like Killing vector\cite{woodardcosmo} that are not available in Minkowsky space time and preclude a spectral representation of correlation functions of the environmental degrees of freedom. In particular there are no kinematic thresholds and quanta of a field can decay in quanta of the \emph{same} field\cite{decayds,akhmedov}, or of a \emph{heavier} field  suggesting that the effective field theory description obtained upon tracing over degrees of freedom with mass scales much larger than the Hubble scale during inflation will always be dissipative and non-unitary. The results of this study will be reported elsewhere.

 \acknowledgements  The author  thanks the N.S.F. for partial
support through grant PHY-1202227.

\appendix

\section{General fluctuation dissipation relation:}\label{app:fd}
Because the initial density matrix $\rho_\chi(0)=e^{-\beta H_{0\chi}}$ is translational invariant and $\mathcal{O}(\vx,t) = e^{iH_{0\chi}t}\,\mathcal{O}(\vx,0)\,e^{-iH_{0\chi}t}$ we can write
\bea  G^>(\vx-\vx';t-t')  & = &  \langle \mathcal{O}(\vx,t)\mathcal{O}(\vx',t') \rangle = \int \frac{d^4k}{(2\pi)^4}~ \rho^>(\vk,k_0) e^{-ik_0(t-t')}\,e^{i\vk\cdot(\vx-\vx')} \label{Ggfd} \\
G^<(\vx-\vx';t-t')  & = &  \langle \mathcal{O}(\vx',t')\mathcal{O}(\vx,t) \rangle = \int \frac{d^4k}{(2\pi)^4}~ \rho^<(\vk,k_0) e^{-ik_0(t-t')}\,e^{i\vk\cdot(\vx-\vx')} \,. \label{Glfd} \eea Introducing a complete set of eigenstates of $H_{0\chi}$ it follows that
\begin{eqnarray}
\rho^>(k_0,\vk) & = &  \frac{2\pi}{\mathrm{Tr}\rho_\chi(0)}~
\sum_{m,n}e^{-\beta E_n}
\langle n| {\cal O}_{\vec k}(0) |m \rangle \langle m| {\cal O}_{-\vec k}(0) |n \rangle \, \delta(k_0-(E_n-E_m)) \label{siggreat} \\
\rho^<(k_0,\vk) & = &  \frac{2\pi}{\mathrm{Tr}\rho_\chi(0)}~
\sum_{m,n} e^{-\beta E_n}
 \langle n| {\cal O}_{-\vec k}(0) |m \rangle \langle m| {\cal O}_{\vec k}(0) |n
 \rangle \, \delta(k_0-(E_m-E_n))
 \label{sigless}
\end{eqnarray} where $\mathcal{O}_{\vk}(0)~;~\mathcal{O}_{-\vk}(0)$ is the spatial Fourier transform of the (composite) operators $\mathcal{O}(\vx,t=0)~;~\mathcal{O}(\vx',t'=0)$ respectively. Upon relabelling
$m \leftrightarrow n$ in the sum in the definition (\ref{sigless})
we find the Kubo-Martin-Schwinger relation\cite{kapusta,lebellac,kms,dasbuk,das2}

\begin{equation}
\rho^<(k_0,k)  = \rho^>(-k_0,k) = e^{-\beta k_0}
\rho^>(k_0,k) \label{KMS}
\end{equation}

\noindent where we have used parity and rotational invariance in
the second line above to assume that the spectral functions only
depend of the absolute value of the momentum. The spectral density is defined as
\be \rho(k_0,k) = \rho^>(k_0,k)-\rho^<(k_0,k) = \rho^>(k_0,k)\big[ 1-e^{-\beta k_0}\big] \label{specOs}\ee
therefore
\be  \rho^>(k_0,k) = \rho(k_0,k)~\big[1+n(k_0)\big]~~;~~\rho^<(k_0,k) = \rho(k_0,k)~ n(k_0) \,. \label{relas}\ee Furthermore, from the first equality in (\ref{KMS}) it follows that
\be \rho(-k_0,k)= - \rho(k_0,k) \,. \label{oddros}\ee

In terms of the spectral densities we find
\be  \Big[G^>(\vx-\vx';t-t')-G^<(\vx-\vx';t-t')\Big]   = \int \frac{d^4k}{(2\pi)^4}\,\rho(k_0,k)  e^{-ik_0(t-t')}\,e^{i\vk\cdot(\vx-\vx')} \label{Godd} \ee
\bea   \Big[G^>(\vx-\vx';t-t')+G^<(\vx-\vx';t-t')\Big] & \equiv &  \int \frac{d^3k}{(2\pi)^3}~\mathcal{K}(k,t-t') e^{i\vk\cdot(\vx-\vx')} \nonumber \\ & = &  \int \frac{d^4k}{(2\pi)^4}\,\widetilde{\mathcal{K}}(k_0,k)  e^{-ik_0(t-t')}\,e^{i\vk\cdot(\vx-\vx')} \label{Geven} \eea where
\be \widetilde{\mathcal{K}}(k_0,k) =  \rho(k_0,k)\,\mathrm{coth\big[ \frac{\beta k_0}{2}\big]}\,.\label{kernelK}\ee

This is the general form of the fluctuation dissipation relation, with $\mathcal{K}$ proportional to the kernel for the noise correlation function (see eqn.(\ref{kernelkappa})). Note that $\rho(k_0,k)$ is \emph{odd}   whereas $\widetilde{\mathcal{K}}(k_0,k)$ is \emph{even} in $k_0$.

\section{Spectral densities:}\label{sec:specdens}

\subsection{\textbf{Case a):} single $\chi$ field of mass $M$ }

The finite temperature correlation function for a single $\chi$ field is
\be G^>(\vx - \vx';t-t')  =   \langle \chi(\vx,t)\chi(\vx',t')\rangle = \int \frac{d^3k}{(2\pi)^3}\Bigg\{ \frac{\big[1+n(w_k)\big]}{2w_k}~e^{-iw_k(t-t')}\,e^{i\vk\cdot(\vx-\vx')}+ \frac{ n(w_k)}{2w_k}~e^{iw_k(t-t')}\,e^{-i\vk\cdot(\vx-\vx')}\Bigg\} \label{G1chi} \ee  where $w_k = \sqrt{k^2+M^2}$. Relabelling $\vk \rightarrow -\vk$ in the second term (\ref{G1chi}) can be written as
\be G^>(\vx - \vx';t-t')  = \int \frac{d^4k}{(2\pi)^4}~ \rho^>(k_0,k) ~e^{-ik_0(t-t')}\,e^{i\vk\cdot(\vx-\vx')} \label{grhog}\ee where
\be \rho^>(k_0,k) = \frac{\pi}{w_k}\Big[\delta(k_0-w_k)-\delta(k_0+w_k) \Big]\,\big[1+n(k_0)\big]\,.\label{rhogreat1chi}\ee In a similar manner we find
\be G^<(\vx - \vx' ;t-t')  =   \langle \chi(\vx',t')\chi(\vx,t)\rangle = \int \frac{d^3k}{(2\pi)^3}\Bigg\{ \frac{\big[1+n(w_k)\big]}{2w_k}~e^{iw_k(t-t')}\,e^{-i\vk\cdot(\vx-\vx')}+ \frac{ n(w_k)}{2w_k}~e^{-iw_k(t-t')}\,e^{i\vk\cdot(\vx-\vx')}\Bigg\}\,, \label{G1chiles} \ee which is written as
\be G^<(\vx - \vx';t-t')  = \int \frac{d^4k}{(2\pi)^4}~ \rho^<(k_0,k) ~e^{-ik_0(t-t')}\,e^{i\vk\cdot(\vx-\vx')} \label{grhol}\ee where
\be \rho^>(k_0,k) = \frac{\pi}{w_k}\Big[\delta(k_0-w_k)-\delta(k_0+w_k) \Big]\, n(k_0)\,.\label{rholes1chi}\ee The spectral density
\be \rho(k_0,k) = \rho^>(k_0,k)-\rho^<(k_0,k)= \frac{\pi}{w_k}\Big[\delta(k_0-w_k)-\delta(k_0+w_k) \Big]\,. \label{specdens1chi}\ee

\subsection{\textbf{Case b):} $\chi_1;\chi_2~;~M_1>M_2$}
Defining in four vector notation $K \equiv (k_0, \vk)$
\be \langle \chi_j(x)\chi_j(y)\rangle  =
\int \frac{d^4K}{(2\pi)^4} ~\Delta_j(K) \,e^{-iK(x-y)}~~;~~j=1,2   \label{Deltas}\ee with
\be \Delta_j(k_0,k) = \frac{\pi}{w^j_k}\Bigg[(1+n(w^{(j)}_k))~\delta(k_0-w^{(j)}_k)+n(w^{(j)}_k)~\delta(k_0+w^{(j)}_k) \Bigg] ~~;~~w^{(j)}_k=
\sqrt{k^2+M^2_j} \label{deltajs}  \ee
With $\mathcal{O}(x) = \chi_1(x)\chi_2(x) $ it follows that
\bea G^>(x-y) & = &  \langle \mathcal{O}(x) \mathcal{O}(y)\rangle = \langle \chi_1(x)\chi_1(y)\rangle\,\langle \chi_2(x)\chi_2(y)\rangle  =  \int \frac{d^4K}{(2\pi)^4}\int \frac{d^4P}{(2\pi)^4}\Delta_1(K)\,\Delta_2(P)\,e^{-i(K+P)(x-y)} \nonumber \\ & \equiv &  \int \frac{d^3q dq_0}{(2\pi)^4} \,\rho^>(q_0,q)\,e^{-iq_0(t-t')}\,e^{i\vq\cdot(\vx-\vy)} \label{gg2chis}\eea from which we obtain
\be \rho^>(q_0,q) = \int \frac{d^3k dk_0}{(2\pi)^4}\, \Delta_1(k_0,k)\,\Delta_2(q_0-k_0, |\vq-\vk| ) \,.\label{rogt}\ee Using (\ref{deltajs}) we find
\bea \rho^>(q_0,q) & = &  \frac{\pi}{2}\int \frac{d^3k}{(2\pi)^3}\,\frac{1}{w^{(1)}_k\,w^{(2)}_{p}}\Bigg\{(1+n_1)(1+n_2)\,\delta(q_0-w^{(1)}_k-w^{(2)}_p) \nonumber \\ &+&
 (1+n_1)n_2 \,\delta(q_0-w^{(1)}_k+w^{(2)}_p)+(1+n_2)n_1 \,\delta(q_0+w^{(1)}_k-w^{(2)}_p)+  n_1 n_2\,\delta(q_0+w^{(1)}_k+w^{(2)}_p)\Bigg\} \label{rho2gt} \eea where (after relabelling $\vk \rightarrow -\vk$)
 \be n_1= n(w^{(1)}_k) ~~;~~ n_2= n(w^{(2)}_p)~~;~~ p = |\vk+\vq| \,.\label{n1n2defs}\ee From the general relation
 \be \rho^<(q_0,q) = \rho^>(-q_0,q) \label{relaqs}\ee obtained above (see eqn. (\ref{KMS})) and $\rho(q_0,q) = \rho^>(q_0,q)-\rho^<(q_0,q)$, we finally find
 \bea \rho(q_0,q)   & = &  \frac{\pi}{2}\int \frac{d^3k}{(2\pi)^3}\,\frac{1}{w^{(1)}_k\,w^{(2)}_{p}}\Bigg\{(1+n_1+n_2)\,\Big[\delta(q_0-w^{(1)}_k-w^{(2)}_p) - \delta(q_0+w^1_k+w^2_p)\Big]\nonumber \\ &+&
 (n_2-n_1) \,\Big[\delta(q_0-w^{(1)}_k+w^{(2)}_p)-\delta(q_0+w^{(1)}_k-w^{(2)}_p)\Big]\Bigg\} \,. \label{rho2chis1} \eea

In the first line in (\ref{rho2chis}) the terms with $n_2$ can be simplified by relabelling $\vk \rightarrow -\vk -\vq$ in the $k$ integral, with $w^{(1)}_k \rightarrow w^{(1)}_p~;~w^{(2)}_p \rightarrow w^{(2)}_k$, then the term proportional to $n_2$ in the first line becomes of the same form as that for $n_1$ but with the replacement $M_1 \leftrightarrow M_2$. In the second line a similar relabelling makes the term proportional to $n_2$ similar to that of $n_1$ upon $M_1 \leftrightarrow M_2$. Therefore
\be \rho(q_0,q) = \rho^{(0)}(q_0,q)+\Big(\rho^{(I)}(q_0,q)+ M_1 \leftrightarrow M_2 \Big)+\Big(\rho^{(II)}(q_0,q)+ M_1 \leftrightarrow M_2 \Big)\,\label{simplerho}\ee where

\begin{eqnarray}
\rho^{(0)}(q_0,q)&=&\frac{1}{16 \pi^2}\, \mathrm{sign}(q_0) \int
\frac{d^3 k}{w^{(1)}_k \,w^{(2)}_p}\,
\delta(|q_0|-w^{(1)}_k -w^{(2)}_p),
\\
\rho^{(I)}(q_0,q)&=&\frac{1}{16 \pi^2}\, \mathrm{sign}(q_0) \int
\frac{d^3 k}{w^{(1)}_k \,w^{(2)}_p}\,n(w^{(1)}_k)\,
\delta(|q_0|-w^{(1)}_k -w^{(2)}_p),
\\
\rho^{(II)}(q_0,q)&=&\frac{1}{16 \pi^2}\, \mathrm{sign}(q_0) \int
\frac{d^3 k}{w^{(1)}_k \,w^{(2)}_p}\,n(w^{(1)}_k)\,
\delta(|q_0|+w^{(1)}_k -w^{(2)}_p),
\end{eqnarray}

Obviously, $\rho^{(0)}(q_0,q$ represents the zero temperature
contribution.  Let
$\omega\equiv \omega_{k}^{(1)}$ and
$z=\omega_{\vec{q}+\vec{k}}^{(2)}$, with
\be d^3k = k^2 dk d(\cos(\theta)) ~~;~~ \frac{d(\cos(\theta))}{w^{(2)}_{\vq+\vk}} = \frac{dz}{kq} \label{changevars}\ee

Then, we have

\bea
\rho^{(0)}(q_0,q)&=&\frac{1}{16 \pi^2}\, \mathrm{sign}(q_0)\int_{M_{1}}^{\infty} \,
d\omega
\int_{z^{-}}^{z^{+}}\delta(\,|q_0|-\omega-z\,)\,dz \label{rho0cal}\\
\rho^{(I)}(q_0,q)&=&\frac{1}{16 \pi^2}\, \mathrm{sign}(q_0)\int_{M_{1}}^{\infty}  n(\omega)  \,
d\omega
\int_{z^{-}}^{z^{+}}\delta(\,|q_0|-\omega-z\,)\,dz \label{rho1cal} \\
\rho^{(II)}(q_0,q)&=&\frac{1}{16 \pi^2}\, \mathrm{sign}(q_0)\int_{M_{1}}^{\infty}  n(\omega)  \,
d\omega
\int_{z^{-}}^{z^{+}}\delta(\,|q_0|+\omega-z\,)\,dz \label{rho2cal}
\,,\eea

where

\be
z^{\pm} = \sqrt{(k\pm q)^{2}+M_{2}^{2}} \\
 =  \sqrt{\omega^{2}+q^2 \pm 2q
\sqrt{\omega^{2}-M_{1}^{2}}-(M_{1}^{2}-M_{2}^{2})}.
\ee

Without loss of generality we   assume that $M_{1}>M_{2}$.

For the integrals in (\ref{rho0cal},\ref{rho1cal}) to be non-vanishing, it must be  that

\begin{equation}
z^{-}<|q_0|-w^{(1)}_{k}<z^{+}. \label{conditionI}
\end{equation}

A simple analysis shows that for large $q_0$ the curve $|q_0|-\omega^{(1)}_{k}$ intersects both $z^{\pm}(k)$, as $|q_0|$ diminishes, both intersections occur with $z^-(k)$ until they coalesce, leading to the condition $Q^2= q^2_0 - q^2 > (M_1+M_2)^2$. We find
\be \rho^{(0)}(q_0,q) = \frac{\mathrm{sign}(q_0)}{8\pi Q^2}\,\Theta(Q^2-(M_1+M_2)^2 )\,\Bigg\{\Big[Q^2-(M_1-M_2)^2 \Big]~\Big[Q^2-(M_1+M_2)^2 \Big]  \Bigg\}^{\frac{1}{2}}\,, \label{rho0fina} \ee
 \be \rho^{(I)}(q_0,q) = \frac{\mathrm{sign}(q_0)}{8\pi \beta q}\,\Theta(Q^2-(M_1+M_2)^2 )\,\ln\Bigg[ \frac{1-e^{-\beta w_+}}{1-e^{-\beta w_-}} \Bigg] \,, \label{rho1fina} \ee where
 \bea && w_{\pm}(q_0,q) = \frac{1}{2Q^2} \Bigg\{|q_0|\, \alpha \pm q\,\sqrt{\alpha^2-4Q^2M^2_1} \Bigg\} \label{wpm} \\
 && \alpha = Q^2 + M^2_1-M^2_2 ~~;~~ \alpha^2-4Q^2M^2_1 = \Big[Q^2-(M_1-M_2)^2 \Big]~\Big[Q^2-(M_1+M_2)^2 \Big]\,. \label{alfas}\eea

The analysis for (\ref{rho2cal}) follows the same steps, now the integral with the delta function is non-vanishing for
\be z^-(k) \leq |q_0|+w^{(1)}_k \leq z^+(k) \,.\label{intro2}\ee There are two different cases: \textbf{i):} $|q_0| < q$ (or $Q^2 <0$) or \textbf{ii):} $|q_0| > q$. In case \textbf{i):} there is intersection only with $z^-(k)$ and the
range of integration in $\omega$ is
\be \xi \leq \omega \leq \infty \,,  \label{rangeneg}\ee where
\be \xi(q_0,q) = \frac{1}{2|Q^2|}\Bigg\{|q_0| \, \alpha + q \sqrt{\alpha^2+4|Q^2|M^2_1}\Bigg\}\,. \label{xis}\ee
In case \textbf{ii):} there are two intersections for $0< Q^2 < (M_1-M_2)^2$ and no intersections for $Q^2 > (M_1-M_2)^2$, combining the two cases we find
\be  \rho^{(II)}(q_0,q) =  \frac{\mathrm{sign}(q_0)}{8\pi \beta q}\Bigg\{\Theta(-Q^2)\ln\Big[ \frac{1}{1-e^{-\beta \xi(q_0,q)}}\Big] -   \Theta(Q^2)\,\Theta((M_1-M_2)^2-Q^2)\,\ln\Bigg[ \frac{1-e^{-\beta w_+}}{1-e^{-\beta w_-}} \Bigg] \Bigg\}\ee

We summarize this result as
\bea \rho(q_0,q;T) & = &  \rho_{LD}(q_0,q;T)_{LD}\,\Theta(-Q^2)+ \rho_D(q_0,q;T)\,\Theta((M_1-M_2)^2-Q^2)\,\Theta(Q^2)\nonumber \\ & + & \rho_{2\chi}(q_0,q;T)\,\Theta(Q^2-(M_1+M_2)^2) ~~;~~Q^2=q^2_0 - q^2 \label{rho2chisapp}\eea
where
\be  \rho_{LD}(q_0,q;T)_{LD} =  \frac{\mathrm{sign}(q_0)}{8\pi \beta q}\Bigg\{\ln\Big[ \frac{1}{1-e^{-\beta \xi(q_0,q)}}\Big]  + M_1 \leftrightarrow  M_2 \Bigg\}\label{rhold}\ee
\be \rho_D(q_0,q;T) = - \frac{\mathrm{sign}(q_0)}{8\pi \beta q}\Bigg\{\ln\Bigg[ \frac{1-e^{-\beta w_+}}{1-e^{-\beta w_-}}\Bigg]+ M_1 \leftrightarrow  M_2    \Bigg\}\label{rhode}\ee
\bea \rho_{2\chi}(q_0,q;T) & = &   \frac{\mathrm{sign}(q_0)}{8\pi Q^2}\Bigg\{ \Big[Q^2-(M_1-M_2)^2 \Big]~\Big[Q^2-(M_1+M_2)^2 \Big]  \Bigg\}^{\frac{1}{2}} \nonumber \\ & + & \frac{\mathrm{sign}(q_0)}{8\pi\beta q} \Bigg\{\ln\Bigg[ \frac{1-e^{-\beta w_+}}{1-e^{-\beta w_-}} \Bigg]+ M_1 \leftrightarrow  M_2\Bigg\} \label{rho2chitos} \eea

\end{document}